\documentclass{aa} 
\usepackage{graphicx}
\usepackage{natbib}
\usepackage{hyperref}
\usepackage{xcolor}

\bibpunct{(}{)}{;}{a}{}{,}
\usepackage{txfonts}
\begin{document} 

   \title{The effect of pressure-anisotropy-driven kinetic instabilities on magnetic field amplification in galaxy clusters}

   \titlerunning{Magnetic field amplification in galaxy clusters}
   \authorrunning{Rappaz \& Schober}

   \author{Y. Rappaz
          \inst{1}
          \and
          J. Schober\inst{1}
          }

   \institute{Institute of Physics, Laboratory of Astrophysics, \'Ecole Polytechnique F\'ed\'erale de Lausanne (EPFL), 1290 Sauverny, Switzerland\\
              \email{yoan.rappaz@epfl.ch}}

  \abstract
    {The intracluster medium (ICM) is the 
    low-density diffuse gas that fills 
    the space between galaxies within galaxy clusters. 
    It is primarily composed of magnetized plasma, which reaches virial temperatures 
    of up to $10^8~\mathrm{K}$, probably due to mergers of subhalos.    
    Under these conditions, the plasma is weakly collisional and therefore has
    an anisotropic pressure tensor with 
    respect to the local direction of the magnetic field. 
    This triggers very fast, Larmor-scale, pressure-anisotropy-driven kinetic instabilities that alter magnetic field amplification.}
    {We study magnetic field amplification through a turbulent small-scale dynamo, including the effects of the kinetic instabilities, during the evolution of a typical massive galaxy cluster. 
    A specific aim of this work is to establish a redshift limit from 
    which a dynamo has to start to amplify the magnetic field up to 
    equipartition with the turbulent velocity field at redshift $z=0$.}
    {We implemented one-dimensional radial profiles for various plasma quantities for
    merger trees generated with the Modified GALFORM algorithm. We assume that turbulence is driven by successive mergers of dark matter halos and
    construct effective models for the Reynolds number $\mathrm{Re}_{\mathrm{eff}}$ dependence
    on the magnetic field in three different magnetization regimes 
    (unmagnetized, magnetized ``kinetic'' and ``fluid''), 
    including the effects of kinetic instabilities.
    The magnetic field growth rate is calculated for the different $\mathrm{Re}_{\mathrm{eff}}$ models.}
    {The model results in a higher magnetic field growth rate at higher redshift. 
    For all scenarios considered in this study, to reach equipartition at $z=0$, it is sufficient for the amplification of the magnetic field to start at redshift $z_{\mathrm{start}} \approx 1.5$ and above.
    The time to reach equipartition can be significantly shorter, in cases with systematically smaller turbulent forcing scales, and for the highest $\mathrm{Re}_{\mathrm{eff}}$ models.
    }
    {The origin of magnetic fields in the weakly collisional ICM can be explained
    by the small-scale turbulent dynamo, provided that the dynamo 
    process starts beyond a given redshift. 
    Merger trees are useful tools for studying
    the evolution of magnetic fields in weakly collisional plasmas,
    and could also be used to constrain the different 
    stages of the dynamo that potentially could be observed by future radio telescopes.}


\maketitle

% *********************************************************************
% *********************************************************************
% *********************************************************************
\section{Introduction}
Magnetic fields are observed in a wide range of systems throughout the Universe.
They are present in stars \citep[e.g.][]{hubrig_2011_bfields_massive_stars}, including the Sun \citep[e.g.][]{Sun_mean_magfield_Scherrer1977,sunspots_magfields}, and the interstellar 
medium of various types of galaxies in the local Universe \citep[e.g.][]{bfields_galaxies_beck_2013} and at intermediate redshifts \citep[e.g.][]{widrow_2002_gal_bf, bernet_2008_strong_bfields_gal}.
Radio observations even reveal the existence of magnetic fields of a few $\mu\mathrm{G}$ in the intracluster medium (ICM) 
\citep[e.g.][]{vogt_2003_abell_400_2634_hydra, bionafede_2010_rot_measure_coma}. 
This corresponds to a magnetic energy density that is close to 
the energy density of turbulent motions. 
Incidentally, magnetic fields of the order of a few $\mu\mathrm{G}$ have been observed in diffuse radio emission in galaxy clusters up to redshift $z \simeq 0.7$ \citep{di_gennaro_2020}.
The origin of such strong magnetic fields, in the largest gravitationally bound objects of the Universe, galaxy clusters, remains unclear. 

The seeds for the ICM magnetic field could have already been generated in the very early Universe, as predicted from certain models of inflation \citep[e.g.][]{fujita_2015_magnetogenesis_axion, adshead_2016_magnetogenesis_axion, talebian_2020_infl_magnetogenesis} and 
cosmological phase transitions \citep{quashnock_1989_QCD_bfields,tornkvist_1998_electroweak_bfield,ellis_2019_phase_trans_bfields}.
Cosmological seed magnetic fields could have been further amplified by a macroscopic quantum effect called the chiral dynamo \citep[e.g.][]{joyce_1997_primordial_bfields, Brandenburg_2017_chiral_effect_early_universe, schober_2022_chiral_anomaly}. 
Due to baryon overdensity, the Biermann battery term \citep{biermann_1950} provides another mechanism for seed field generation before recombination \citep[e.g.][]{naoz_narayan_2013}.
However, observations of the cosmic microwave background provide 
strict upper limits for the comoving field strength of
cosmological seed fields between $10^{-9}$ \citep{planck_2016} 
and $10^{-11}~\mathrm{G}$ \citep{jedamzik_2019_CMB_limit}
on length scales above approximately one $\mathrm{Mpc}$.
Alternatively, seed fields could be generated through astrophysical mechanisms, like the Biermann battery
during structure formation \citep[e.g.][]{biermann_kulsrud_1997, biermann_ryu_1998} or reionization \citep[e.g][]{biermann_subramanian_1994, gnedin_2000_reionization}.
Yet astrophysical magnetogenesis produces magnetic fields that are too weak to explain the observed values in the ICM. Hence, an efficient amplification process is needed.

One commonly discussed candidate for magnetic field amplification is the 
small-scale turbulent dynamo
\citep{Schekochihin_2006_magfield_gclust, Miniati_2015_ICM_dynamo, Domínguez-Fernandez_2019_cluster}.
This is a magnetohydrodynamical (MHD) phenomenon, in which 
turbulence stretches, twists, and folds the magnetic field on small spatial scales. 
In the kinematic stage, the magnetic energy spectrum exhibits
a Kazantsev-like power law \citep{kazantsev_1968} with a peak at the resistive scale $k_{\eta}$ 
\citep{Schekochihin_2004_small_scale_dynamo, brandenburg2023_diss_ssd}.
For typical plasma parameters of the ICM, 
this corresponds to $k_{\eta}\approx (10^4~\mathrm{km})^{-1}$ \citep{Schekochihin_2006_magfield_gclust}.
However, observations of the Faraday rotation measure in the ICM 
reveal a typical reversal scale of the magnetic field of a few kiloparsecs \citep[e.g.][for the Coma cluster]{bionafede_2010_rot_measure_coma}. 
This could indicate that the small-scale dynamo is in the non-linear evolutionary stage, where magnetic energy shifts to larger spatial scales \citep{scheko_2002_nonlin_MHD_dnyamo, beresnyak_2012_universal_ssd, schleicher_2013_ssd}.
Whether or not this scenario can explain the 
existence of $\mu\mathrm{G}$ magnetic fields in galaxy clusters, depends
on the properties of ICM turbulence across cosmic times.

In hierarchical structure formation, 
galaxy clusters have gradually formed through a series of dark
matter halo mergers and gas accretion \citep[e.g.][]{mo_galaxy_formation_notebook}. 
Such mergers are some of the most 
energetic events in the Universe 
and can drive turbulence. 
The hierarchical assembly of dark matter halos leading to the 
formation of galaxy clusters has been extensively
studied, 
both analytically through the extended 
Press-Schechter theory \citep{press_schechter_model_1974, cole2000_galform} 
and via cosmological simulations; e.g.~Millennium \citep{springel_2005_millenium} 
and IllustrisTNG \citep[e.g.][]{marinacci_illustris_2018}. 
Such tools can be employed to study the 
evolution of turbulence and therefore magnetic 
field amplification in the ICM.
In fact, the growth rate of the classical 
small-scale dynamo in its kinematic stage increases with 
increasing Reynolds number $\mathrm{Re}$ 
\citep{kazantsev_1968}, 
which measures the ratio between inertial and viscous forces.
While in galaxies $\mathrm{Re}$ reaches values of
approximately $10^{10}$ \citep[e.g.][]{schober_2013_youngal} and therefore allows to explain the generation of the strong 
(random component) of galactic magnetic fields by a small-scale 
dynamo, 
it is estimated to be only on the order of a 
few tens in the ICM of local clusters \citep[assuming the 
standard Spitzer viscosity; e.g.][and references therein]{Schekochihin_2006_magfield_gclust}. 
With such a low value of $\mathrm{Re}$, amplifying the 
seed magnetic fields to equipartition with turbulent gas motions 
through a classical small-scale dynamo might not be possible.

Another complication of the classic small-scale dynamo
scenario arises from the fact that MHD theory might not be suitable for describing 
the ICM. 
The applicability of MHD depends on the collisionality of the plasma. 
The ion-ion collision frequency
$\nu_{ii}$ is inversely proportional to the plasma temperature $T$
as 
$\nu_{ii} \propto T^{-3/2}$ \citep{fitzpatrick2014plasma}. 
With typical virial temperatures of $10^7$ to 
$10^8~\mathrm{K}$ generated by mergers of dark matter halos, 
$\nu_{ii}^{-1}/t_{\mathrm{clust}}\approx 10^{-2}$ 
\citep[e.g.][]{ann_ewv_galaxy_clusters_formation}, where $t_{\mathrm{clust}}$ is the typical dynamical time of a cluster.
This suggests that the ICM plasma is weakly collisional.
In weakly collisional plasmas (WCPs), the first magnetic moment 
of a particle, $\mu \equiv mv_{\perp}^{2}/(2B)$, is an adiabatic 
invariant, meaning that whenever a change in the magnetic field $B$
is produced, the perpendicular velocity $v_{\perp}$ of a particle with mass $m$
adjusts in order to conserve $\mu$.
This produces a bias in the thermal motions of the particles, leading to pressure anisotropies along and perpendicular to the local direction of the magnetic field.
The main issue is that, in a $\mu$-conserving environment, dynamo action is impossible.
Indeed, when the magnetic field grows, a corresponding change in the anisotropic 
temperature should occur, 
but there is not enough thermal energy in the plasma 
to be redistributed that way \citep{helander_strumik_schekochihin_2016}.
Besides, pressure anisotropies trigger microscale, 
fast-growing kinetic instabilities that will break 
the $\mu$-invariance, therefore allowing the dynamo process to kick in.
The most commonly encountered instabilities in 
literature are the mirror and the firehose 
ones\footnote{The mirror instability occurs when a magnetic bottle 
grows in amplitude and is destabilized by an excess of perpendicular 
pressure over magnetic pressure.
The firehose instability is developed when the restoring force of
propagating Alfvénic fluctuations is undermined by an opposing 
viscous stress, making those fluctuations grow exponentially.}
\citep[e.g.][]{rosenbluth_1956_pinch, parker_1958_aniso_gas, vedenov_sagdeev_1958, gary_1992, southwood_kivelson_1993, hellinge_2007, kunz_scheko_stone_2014, melville_2016}.
Therefore, in WCPs, discarding kinetic effects 
from macroscale dynamics is no longer possible, which constitutes an enormous theoretical and numerical challenge.

In the past years, an extensive amount of simulations and theoretical modeling 
of WCPs
have been conducted, for different magnetization states. 
For instance, \citet{santos-lima-2014} performed simulations of 
the turbulent dynamo based on the double-adiabatic Chew-Goldberger-Low equations
\citep{chew_goldberger_low_1956_closure}, in which they implemented an 
anomalous collisionality aiming to mimic the particle-scattering 
produced by firehose and mirror instabilities. 
Furthermore, \citet{Rincon_2016_colless_dynamo} 
used low-resolution hybrid-kinetic simulations of a turbulent 
dynamo in a collisionless unmagnetized plasma to demonstrate that 
a dynamo effect is effectively possible in such a system. 
Hybrid particle-in-cell simulations of the turbulent dynamo 
in the ``kinematic'' and non-linear phases, as well as in the saturated state of a collisionless magnetized plasma
have been performed by \citet{st-onge_2018_hybrid}, 
where they calculated the effective collisionality 
produced by the microscale kinetic instabilities. 
Finally, \citet{st-onge_kunz_squire_scheko_20202} performed 
extensive simulations of a WCP, based on
MHD equations supplemented by a field-parallel Braginskii viscous stress. 
A significant result from the two latter studies is that the parallel viscosity 
``knows'' about the direction and strength of the magnetic field, 
which basically leads to a dependency of the Reynolds number on the magnetic field,
meaning that the Reynolds number $\mathrm{Re}$ is 
replaced by an effective version $\mathrm{Re}_{\mathrm{eff}}(B)$.
In the magnetized ``fluid'' regime 
(corresponding to magnetic field strengths above a few $n\mathrm{G}$), \citet{st-onge_kunz_squire_scheko_20202} argued that the effective Reynolds number scales as 
$\mathrm{Re}_{\mathrm{eff}} \approx \beta^2 \mathcal{M}^4$, where $\beta$ (thermal pressure to magnetic pressure ratio) and $\mathcal{M}$ are the plasma 
parameter and the Mach number, respectively. 
On the other hand, \citet{Rincon_2016_colless_dynamo} found that in 
the unmagnetized (collisionless) regime, the magnetic field evolves 
into a folded structure with a magnetic power spectrum close to 
the high magnetic Prandtl number MHD case. 
However, the form of  this relation in the magnetized 
``kinetic'' regime is still an open question.
Despite advancements in simulations and modeling, investigating 
the dynamics of magnetic fields over cosmological timescales on  
which galaxy clusters form remains a challenging task. 
This is due to the vast range of values of 
$\rho_\mathrm{i}/L_0$ (with $\rho_\mathrm{i}$ being the ion Larmor radius and 
$L_0$ the fluid-scale stretching) that must be covered, going 
from seed-field values (which can be as low as
$10^{-20}~\mathrm{G}$) to values close to equipartition 
in the ICM (approximately $10^{-6}~\mathrm{G}$). 
This range covers both the kinetic and fluid scales, 
making it computationally infeasible at present.

In this paper, we circumvent the difficulties related to the 
resolution of cosmological simulations by using a semi-analytical approach based 
on merger trees. 
These algorithms function by tracing the merger history 
of dark matter halos and are used either as a compact representation 
of the merging/accretion rate of dark matter halos in simulations 
\citep[e.g.][]{tweed_2009_mt_extract_from_sim} or as 
statistical tools to study the formation and evolution of 
cosmic large-scale structures \citep[e.g.][]{dvorkin_Rephaeli_2011, gomez_2022_mt_galaxies}. 
In particular, \citet{cole2000_galform} developed the GALFORM model, which is based on the extended Press-Schechter theory \citep{snyder_1997} 
to study the evolution of galaxy formation.
This model was later extended by \citet{parkinson_2008_mod_galform}, who modified
the conditional mass function from the latter in order to match the 
statistical predictions from the Millenium simulations 
\citep{springel_2005_millenium}. 
Our objective is to generate merger trees based on the Modified 
GALFORM algorithm and study 
the evolution of magnetic fields in galaxy clusters, by implementing
a semi-analytical model of the dynamo based on the results from 
state-of-the-art simulations of WCPs. 

The paper is organized as follows. 
In Sec.~\ref{sec:theory}, we present the global theoretical framework 
of the different magnetization regimes in WCPs, 
along with the physics of the pressure-anisotropy-driven microscale 
instabilities. 
In Sec.~\ref{sec:model}, we present the Modified GALFORM 
algorithm and the construction of our semi-analytical model of 
the turbulent dynamo. 
Our results are discussed in Sec.~\ref{sec:results} and our 
conclusions are presented in Sec.~\ref{sec:conclusions}.

% *********************************************************************
% *********************************************************************
% *********************************************************************
\section{Theory of weakly collisional plasmas}\label{sec:theory}
To understand how initial seed fields are amplified by dynamo processes in WCPs, we must divide the dynamo problem into different magnetization regimes, and deal with them separately.

\subsection{The unmagnetized regime (UMR)}

The UMR corresponds to the situation where the ion cyclotron frequency $\Omega_\mathrm{i}$ is much smaller than any other dynamical frequency $\omega$, and the Larmor radius $\rho_\mathrm{i}$ is much larger than any other  
spatial scale $l_\mathrm{clust}$ of the system. 
In other words, we have 
$\Omega_\mathrm{i} / \omega \ll 1$ and $\rho_\mathrm{i}/l_\mathrm{clust} \gg 1$. 
This corresponds to magnetic field amplitudes of
\begin{equation}\label{eq:unmag_condition}
B \lesssim 10^{-18} \left(\frac{n_e}{10^{-3}~\mathrm{cm}^{-3}}\right) \left(\frac{T}{5~\mathrm{keV}}\right)^{-3/2}~\mathrm{G},
\end{equation}
where $n_e$ is the electron density. Equation \eqref{eq:unmag_condition} indicates that magnetic fields of the order of $10^{-18}~\mathrm{G}$ 
are strong enough to drive the ICM into a magnetized 
regime \citep{st-onge_kunz_squire_scheko_20202}.

Fully kinetic numerical simulations of the Vlasov equation have been conducted by \cite{Rincon_2016_colless_dynamo}, and have shown that in an initially unmagnetized plasma, magnetic field amplification does occur in stochastically driven, non-relativistic flows. Moreover, they have shown that the magnetic field evolves with a power spectrum that resembles its high-Prandtl-number MHD counterpart. This is a solid justification that turbulent folding of the magnetic field is a main process that amplifies magnetic energy during merger events during the formation of a galaxy cluster.

\subsection{The magnetized regime (MR)}
As soon as $\rho_\mathrm{i} \approx l_\mathrm{clust}$ and $\Omega_\mathrm{i} \approx \omega$, 
the system enters the magnetized regime, and the new scaling relations will drastically change the magnetic field amplification problem. Magnetization along with the condition $\rho_{\mathrm{i},\alpha}/\lambda_{\mathrm{mfp, \alpha}} \ll 1$, where $\lambda_{\mathrm{mfp, \alpha}}$ stands for the mean free path of particle species $\alpha$, leads to a pressure tensor $\overline{\boldsymbol{p}}$ with anisotropic components with respect to the local direction of the magnetic field $\boldsymbol{\hat{b}} \equiv \boldsymbol{B}/B$, that can be expressed as:
\begin{equation}\label{eq:pressure_tensor}
 \overline{\boldsymbol{p}} = p_{\perp}\overline{\boldsymbol{I}}+(p_{\parallel}-p_{\perp})\boldsymbol{\hat{b}}\boldsymbol{\hat{b}},
\end{equation}
where $\overline{\boldsymbol{I}}$ is the unity tensor, and $p_{\parallel}$ and $p_{\perp}$ are respectively the pressure along and perpendicular to the local direction of the magnetic field. The pressure anisotropy term $p_{\parallel}-p_{\perp}$ plays a major role in the stability of a WCP, as will be explained below. 
By taking the first velocity moment of the Vlasov equation 
in the guiding-center limit 
\citep{kulsrud_1983_MHD_description}, we obtain the following 
modified Euler momentum equation:
\begin{equation}\label{eq:euler_mod_eq}
m n\frac{d  \boldsymbol{v}}{dt} = -\nabla \left(p_{\perp}+\frac{B^2}{8\pi}\right)+\nabla \cdot \left[ \boldsymbol{\hat{b}}\boldsymbol{\hat{b}} \left(p_{\perp}-p_{\parallel}\right)+\frac{B^2}{4\pi}\right],
\end{equation}
where $m, n$, and $\boldsymbol{v}$ 
respectively are the proton mass, density, and velocity, 
and where $d/dt \equiv \partial_t +(\boldsymbol{v}\cdot \nabla)$ is the Lagrangian derivative along the trajectory of a fluid element.
Equation \eqref{eq:euler_mod_eq} requires an additional closure relation constraining the evolution of the anisotropic pressure.
Considering the guiding-center limit of the Vlasov equation \citep{kulsrud_1983_MHD_description}, we obtain the double-adiabatic equation of state, also known as the Chew-Goldberger-Low (CGL) equations 
\citep{chew_goldberger_low_1956_closure}, written as
\begin{equation}\label{eq:CGL_equations}
\frac{d}{dt}\left(\frac{p_{\perp}}{nB}\right) = \frac{d}{dt}\left(\frac{p_{\parallel}B^2}{n^3}\right) = 0.
\end{equation}
In the limit of the CGL equations, any local change of the magnetic field creates a corresponding local pressure anisotropy \citep{rincon_2019_dynamothies}:
\begin{equation}\label{eq:delta_pressure}
\Delta \equiv \frac{p_{\perp}-p_{\parallel}}{p},
\end{equation}
where $p \equiv (p_{\parallel}+2p_{\perp})/3$. However, $\mu$-conservation constitutes in itself a fundamental problem regarding dynamo effects. 
Indeed, it has been pointed out by \citep{Kulsrud_1997_cosmic_dynamos} 
that any increase in the magnetic field amplitude has 
to cause the same order-of-magnitude increase in the perpendicular thermal energy of the plasma, which is energetically
impossible in 
a high-$\beta$ plasma stirred by subsonic turbulent motions. In other words, magnetic-field amplification is impossible in a $\mu$-conserved, magnetized plasma \citep{helander_strumik_schekochihin_2016, santos-lima-2014}. For a dynamo action to kick in, $\mu$-conservation has to be broken.
In the Braginkskii limit \citep{braginskii_1965}, where the collision frequency is 
large compared to the inverse of the timescale 
on which the magnetic field strength changes, we obtain the following approximation:
\begin{equation}\label{eq:delta_inst}
\Delta \simeq \frac{1}{\nu_{ii}}\frac{1}{B}\frac{dB}{dt}
= \frac{1}{\nu_{ii}} \hat{\boldsymbol{b}}\hat{\boldsymbol{b}}: \nabla \boldsymbol{u}.
\end{equation}
Note that the second equality in the equation above comes from 
the classical ideal induction equation, and the assumption of incompressibility of the flow.
Equation~(\ref{eq:delta_inst}) implies that any change in the magnetic field will automatically create corresponding changes in the pressure anisotropies.
Whenever $\Delta$ grows, the system is driven away from stability, and plasma instabilities are triggered, whose most famous ones are called mirror and firehose \citep{rosenbluth_1956_pinch, CKW_1958_pinch, parker_1958_aniso_gas, vedenov_sagdeev_1958, gary_1992, southwood_kivelson_1993, hellinge_2007, melville_2016}. 
The interested reader is invited to consult the extensive aforementioned literature for more details about those instabilities, as we will only list basic concepts here.
The mirror instability is triggered when $\Delta > 1/\beta$ 
and the firehose one when $\Delta \lesssim -2/\beta$. 
All those instabilities develop on 
scales close to the ion Larmor radius, at a frequency close 
to the gyrofrequency \citep[e.g.][]{melville_2016}.
In other words, they are much faster compared to the fluid-scales dynamics of a WCP. The question is, what is the effect of those instabilities on the global dynamics of the magnetic field?
In other words, which mechanism allows those kinetic-scale instabilities to pin the system at the stability threshold during the non-linear stage?
Two scenarios are found in the literature.
There is a possibility that mirror and firehose fluctuations screen particles from variations of the magnetic-field amplitude.
For instance, in the firehose case, the stretching rate of the magnetic field by turbulent eddies is reduced, which could also be seen as an enhancement of the plasma viscosity \citep{mogavero_scheko_2014}. Another possible effect of the kinetic-scale instabilities is to scatter particles \citep{kunz_scheko_stone_2014}.
This particle-scattering process acts as an effective reduction of the plasma viscosity.
Overall, both processes are observed in simulations \citep{kunz_scheko_stone_2014, riquelme_quataert_verscharen_2105}.
The transition between those two processes is still not clear, although the screening mechanism seems to be predominant over the particle-scattering one for small magnetic field variations $\delta B/B \ll 1$ \citep{rincon_2019_dynamothies}.

Given that effects from pressure-anisotropy-induced kinetic instabilities are much faster compared to the dynamical timescales of the system considered, a common numerical method to prevent the system from crossing the stability thresholds employs the so-called 
``hard wall limiters'' 
\citep{Sharma_2006_hwl} that takes the form
\begin{equation}\label{eq:hard_wall_limiters}
p_{\perp}-p_{\parallel} = 
\begin{cases}
\max \left(-\frac{B^2}{4\pi}, 3\mu_B \hat{\boldsymbol{b}}\hat{\boldsymbol{b}}: \nabla \boldsymbol{u}\right) &\text{firehose limit}\\
\min \left(\frac{B^2}{8\pi}, 3\mu_B \hat{\boldsymbol{b}}\hat{\boldsymbol{b}}: \nabla \boldsymbol{u}\right) &\text{mirror limit}\\
\end{cases}
\end{equation}
with $\mu_B$ being the Braginskii viscosity \citep{braginskii_1965}.
Pinpointing the system at the instability threshold using \eqref{eq:hard_wall_limiters} implies a modified collisionality of the form 
$\nu_{ii, \mathrm{eff}}\propto \beta \hat{\boldsymbol{b}}\hat{\boldsymbol{b}}: \nabla  \boldsymbol{u}$, which leads to an effective Reynolds number that depends on the magnetic field through the 
$\beta$-parameter \citep{st-onge_kunz_squire_scheko_20202} as
\begin{equation}\label{eq:effective_reynolds_number_magfluid}
\mathrm{Re}_{\mathrm{eff}} \approx \beta^2\left(\frac{v_{\mathrm{turb}}}{v_{\mathrm{th},i}}\right)^4.
\end{equation}
However, such limiters can only be applied when $\nu_{ii, \mathrm{eff}} \lesssim \Omega_\mathrm{i}$, which corresponds approximately 
to magnetic field amplitudes above
\begin{equation}\label{eq:magfluid_limit}
B \gtrsim 6 \left(\frac{n_e}{10^{-3}~\mathrm{cm}^{-3}}\right)^{2/5}\left(\frac{T}
{5~\mathrm{keV}}\right)^{1/2}\left(\frac{\mathcal{M}}{0.2}\right)^{3/5}
\left(\frac{L_0}{100~\mathrm{kpc}}\right)^{-1/5}~\mathrm{nG}
\end{equation}

To sum up, in order to model the magnetic field evolution from seed-field values to equipartition-level values observed in low-redshift galaxy clusters, we have to consider three different magnetization regimes \citep{st-onge_kunz_squire_scheko_20202}: 
\begin{enumerate}
    \item Unmagnetized regime, where condition \eqref{eq:unmag_condition} holds,
    \item Magnetized ``kinetic'' regime, 
    in which hard-wall limiters \eqref{eq:hard_wall_limiters} cannot be applied,
    \item Magnetized ``fluid'' regime, 
    where condition \eqref{eq:magfluid_limit} holds, in which pressure-anisotropies are regulated by kinetic instabilities.
\end{enumerate}
Therefore, modeling the evolution of the magnetic field during the entire formation of a galaxy cluster 
requires covering a wide range of values for the ratio of the Larmor radius over the typical length scale of the system, which is way out of current numerical capabilities.  

% *********************************************************************
% *********************************************************************
% *********************************************************************
\section{Model for magnetic field evolution in clusters}\label{sec:model}

\subsection{Merger tree code and general strategy}\label{subsec:general}
In the standard model of cosmology, 
all structures form from initial primordial density fluctuations. 
Merger trees constitute a valuable tool in modeling 
the statistical evolution of dark matter halos. 
We use the output of such tools, i.e.~the history of 
the assembly of various dark matter halos, to model 
the cosmological evolution of plasma parameters 
(temperature, gas density, turbulence nature, etc.). 
This allows us to calculate the growth rate of the 
small-scale dynamo, and therefore enables us to 
study the evolution of the magnetic field over 
cosmological timescales. 
This is currently unfeasible through numerical 
simulations, given that the computational power 
required to model the dynamics of kinetic instabilities 
surpasses our current capabilities.

To generate dark matter merger trees, we use the Modified GALFORM code 
described in \citet{parkinson_2008_mod_galform}. 
This code is a modification of the GALFORM semi-analytical 
model developed by \citet{cole2000_galform}.
The latter describes cosmic structure formation by employing the 
mass function from the extended Press-Schechter theory, also 
implementing star formation feedback, chemical evolution, dust, starburst phases, etc. 
For a more detailed description of the GALFORM algorithm, 
the interested reader is invited to consult the original 
papers or see Appendix \ref{appendix:galform}.
In this study, we use the cosmological parameters obtained from Planck
\citep{planck_2020_cosmology2018}, i.e.~$\Omega_\mathrm{m} = 0.315$,
$\Omega_\Lambda = 0.685$,
and $H_0 = 67.4~\mathrm{km}\ \mathrm{s}^{-1}\mathrm{Mpc}^{-1}$.

We generate merger trees with a final mass of 
$M_{z=0} = 10^{15}~\mathrm{M}_{\odot}$, 
and consider values of the maximum redshift of
$z_{\mathrm{max}} = 1,1.5,2,...,5$. 
We employ a constant sampling in the redshift space 
of $\Delta z = z_{\mathrm{max}}/300$, which easily satisfies 
the condition 
$P \ll 1$, where $P$ is the average number of progenitors in a given mass range defined by \eqref{eq:PS_prob_halos}. 
The mass resolution is set to $M_{\mathrm{res}} = 10^{12}~M_{\odot}$. 
Finally, we generate a total of $N_{\mathrm{tree}} = 10^3$ merger trees for each set of parameters.
In Appendix~\ref{appendix:nb_trees_effect} we discuss the effect 
of the number of merger trees on the evolution of various physical quantities. 
It is shown in Fig.~\ref{appendix:nb_trees_effect} 
that increasing the number of merger trees 
above $N_{\mathrm{tree}} \approx 500$ 
has very little impact on the final curves. 

\subsection{Radial profiles}1
\label{sec:DM_baryons_distrib}

For each $i$-th subhalo $H^{(i)}_{z_{\alpha}}$ at a given redshift $z_{\alpha}$,
we assume that the dark matter distribution, $\rho_{\mathrm{DM}}$, 
is determined by the Navarro-Frenk-White profile 
\citep{navarro_frenk_white_1996}, which is given by 
\begin{equation}
  \rho_{\mathrm{DM}}(r) = \frac{\rho_\mathrm{s}}{x(1+x)^2},
\end{equation}
where $\rho_\mathrm{s}$ is an integration constant and $x \equiv r/r_\mathrm{s}$, where $r$ is the radial coordinate. 
The scale radius $r_\mathrm{s}$ is specified by the concentration 
parameter $c$ as $r_{\mathrm{vir}} \equiv cr_\mathrm{s}$.
Therefore, determining $c$ and $\rho_\mathrm{s}$ for each subhalo 
gives us directly the dark matter profile.
We determine $c$, following a similar approach as used in \citet{dvorkin_Rephaeli_2011}, where  
energy conservation is imposed after multiple mergers or simple matter accretion.
A detailed description of this process is presented in Appendix~\ref{appendix:c_param}.
We further use the result reported by \citet{ostriker_2005_T_and_rho}, 
which states that
the gas density and temperature distributions, assuming a 
hydrostatic equilibrium of a polytropic gas in a dark matter halo, is given by
\begin{equation}\label{eq:dens_baryons}
\rho_\mathrm{g}(r) =
\rho_0 \left[1-\frac{\Lambda}{1+\xi}\left(1-\frac{\ln\left(1+r/r_\mathrm{s}\right)}{r/r_\mathrm{s}}\right)\right]^{\xi}, 
\end{equation}
and
\begin{equation}\label{eq:temp_baryons}
T(r) = T_0 \left[1-\frac{\Lambda}{1+\xi}\left(1-\frac{\ln\left(1+r/r_\mathrm{s}\right)}{r/r_\mathrm{s}}\right)\right],
\end{equation}
where
\begin{equation}
\Lambda \equiv \frac{4\pi G \rho_\mathrm{s} r_\mathrm{s}^2 \mu m_\mathrm{p}}{k_\mathrm{B}T_0},
\end{equation}
with $\mu m_\mathrm{p}$ being the mean molecular weight, 
$\xi \equiv (\Gamma-1)^{-1}$ the polytropic index, 
$m_\mathrm{p}$ the proton mass, and $k_\mathrm{B}$ the Boltzmann constant.
We adopt $\mu = 1/2$ and $\Gamma = 1.2$. 
We first determine the constant $T_0$ by imposing that the 
temperature at the virial radius should equal the virial temperature of a truncated, single isothermal sphere, which yields
\begin{equation}
T(r_{\mathrm{vir}}) \equiv T_{\mathrm{vir}} = 
\frac{\mu m_\mathrm{p}}{k_\mathrm{B}}v_{\mathrm{vir}}^2 \simeq 3.6\cdot 10^5~\mathrm{K} \left(\frac{v_{\mathrm{vir}}}{100~\mathrm{km}/\mathrm{s}}\right)^2,
\end{equation}
where the virial velocity is approximately given by
\begin{equation}
v_{\mathrm{vir}}^2 \approx \frac{GM}{r_{\mathrm{vir}}}.
\end{equation}
Then, the constant $\rho_0$ is calculated by integration:
\begin{equation}
   \int_{0}^{r_{\mathrm{vir}}}4\pi r^2\rho_\mathrm{g}(r)dr = M_\mathrm{b}
\end{equation}
where $M_\mathrm{b}$ is the total mass of baryons in each dark matter subhalo. 
We assume that $M_\mathrm{b}/M_{\mathrm{DM}} = 0.1$ \citep[e.g.][]{mccarthy_2007_baryon_frac_clusters}.

Our model of turbulent velocity is based on the work 
of \citet{shi_2018}. 
They investigated the evolution of turbulence in the intracluster medium after major mergers using a set of non-radiative hydrodynamical cosmological simulations and 
highlighted that the turbulent velocity increases with radius.\footnote{This effect could be caused by the combined effects of density stratification and faster eddy turnover time in the ICM's core.} 
They also analyzed the radial dependence of the turbulent power spectrum, 
and highlighted that the turbulent injection scale is independent of the radius.
Therefore, we assume the turbulent velocity of each subhalo to be
\begin{equation}\label{turb_vel_profile}
v_{\mathrm{turb}}(r) \equiv v_0 \left(1+\frac{r}{r_{\mathrm{vir}}}\right)^{1/2}
\end{equation}
where $v_0$ is an integration constant. 
We calculate the latter making use of the work 
by \citet{vazza_roediger_2012}. The turbulent injection length is assumed to be of the form 
\begin{equation}\label{equ::inject_length}
L_{0} \equiv \frac{1}{\alpha_0}r_{\mathrm{vir}}
\end{equation}
where $\alpha_0$ is a free parameter,
allows us
us to study the effect of various injection scales on our model. 
The different values tested are respectively $\alpha_0 = 5,10,20$ and 
$50$.\footnote{Note that $\alpha_0 \simeq 20$ is the result obtained in \citet{shi_2018}. The last value $\alpha_0 = 50$ is chosen arbitrarily.}

\subsection{Plasma parameters deduced from averaged profiles}\label{subsec:plasma_params}

Now we assign a unique scalar value to each subhalo $H^{(i)}_{z_{\alpha}}$, by averaging 
over the one-dimensional, spherically symmetric, 
radial profiles described in the previous section.
For a given physical quantity $\Phi \in (T, c, \rho_\mathrm{g}, v_\mathrm{turb})$, 
we calculate the following average $\langle \Phi\rangle_{\rho_g}$, weighted by the gas density distribution:
\begin{equation}\label{eq:avg_dens}
\langle \Phi\rangle_{\rho_g} \equiv  \frac{\int_{0}^{r_{\mathrm{vir}}}\Phi(r)\rho_g(r)dr}{ \int_{0}^{r_{\mathrm{vir}}}\rho_g(r)dr}.
\end{equation}
Here, $r_{\mathrm{vir}}$ is the virial radius calculated for each subhalo by the relation
\begin{equation}
M_{\mathrm{vir}} = \frac{4}{3}\pi r_{\mathrm{vir}}^3 \Delta_c \rho_{c},
\end{equation}
where $M_{\mathrm{vir}}$ is the virial mall of the subhalo, $\rho_c$ is the critical density of the Universe at a given redshift, and $\Delta_c$ is the overdensity constant. We adopt $\Delta_c = 200$.
Then, at redshift $z_{\alpha}$, all values resulting from \eqref{eq:avg_dens} are again 
averaged
with the mass of each subhalo. All in all, each merger tree is attributed a one-dimensional, time-evolving profile given by 
\begin{equation}\label{eq:avg_mz}
    \overline{\langle \Phi\rangle_{\rho_g}}(z_{\alpha}) \equiv \frac{\sum_{i \in z_{\alpha}} M_i \langle \Phi\rangle_{\rho_g}}{\sum_{i \in z_{\alpha}} M_i}.
\end{equation}
The index $i \in z_{\alpha}$ indicates that the sum is performed over all subhalos present at redshift $z_{\alpha}$.

This approach results in a more compact representation of the evolution of the different plasma parameters. 
Indeed, we want to establish a global trend, that does not take into account the numerous dynamic phenomena operating within the ICM, such as
active galactic nuclei jets cooling flows or galactic wakes, etc. 
\citep[e.g.][and references therein]{subramanian_2006}. 
It also allows us to bypass the difficult task of implementing a radial profile for the magnetic field in each subhalo.
For instance, we would need to assume how the magnetic energy 
of progenitors is distributed after a merging event in the 
final halo or assume the resulting distribution. 
Finally, we average the $N_{\mathrm{trees}} = 10^3$ distributions of all physical parameters by fitting all values at a given redshift with a Skew-normal distribution.
A detailed explanation of this procedure is given in Appendix~\ref{appendix:skew-normal}.

From the averaged values $\langle \Phi\rangle_{\rho_g}$, we 
deduce more plasma parameters, while considering
a completely ionized plasma made of protons and electrons.
The thermal velocity of particle $\alpha$, where $\alpha$ refers 
either electrons or ions, is given by
\begin{equation}
v_{\mathrm{th},\alpha} \equiv \sqrt{\frac{2k_\mathrm{B}T}{m_{\alpha}}}.
\end{equation}

The ion-ion collision frequency is calculated assuming momentum exchange only between protons, because of the ratio of 
one thousand 
between proton and electron masses. The derivation can be found, for instance, in \citet{fitzpatrick2014plasma}, and leads to
\begin{equation}
\nu_{ii} = \frac{8}{3}\left(\frac{m_e}{m_i}\right)^{1/2}\frac{\pi^{3/2}e^4n_e\log \Lambda}{\sqrt{m_e}}     \frac{1}{\left(k_\mathrm{B}T\right)^{3/2}}.
\end{equation}
The parallel viscosity which only damps motions parallel to the local direction of the magnetic field, derived in \citet{braginskii_1965}, is given by
\begin{equation}
\mu_{\parallel} \equiv v_{\mathrm{th},\alpha}\lambda_{\mathrm{mfp},\alpha}.
\end{equation}
The Reynolds number is defined as 
\begin{equation}\label{eq:classical_reynolds_number}
\mathrm{Re} \equiv \frac{v_{\mathrm{turb}}L_0}{\mu_{\parallel}},
\end{equation}
where $L_0$ is the injection scale of turbulence. 
Finally, the equipartition field strength is given by
\begin{equation}
B_{\mathrm{equ}} = \sqrt{4\pi \rho_\mathrm{g} v_{\mathrm{turb}}^2}.
\end{equation}

\subsection{Magnetic fields}\label{sec:model_magfields}
It has been mentioned in Sec.~\ref{sec:theory}, that in the magnetized 
``fluid'' regime, 
pressure anisotropies could be well regulated by kinetic-scale instabilities, 
like the firehose and mirror instabilities. 
This leads to a relation 
$\mathrm{Re}_{\mathrm{eff}} = \mathrm{Re}_{\mathrm{eff}}(B)$, namely, that given by Eq.~(\ref{eq:effective_reynolds_number_magfluid}).
On the other hand, the latter does not hold in the magnetized 
``kinetic'' regime.
The instabilities do not scatter particles fast enough to keep the pressure anisotropies near marginal stability. 
Therefore the exact dependency of the effective Reynolds 
number on the magnetic field in this regime
is still an open question.

\begin{figure}
    \centering
    \includegraphics[width = \columnwidth]{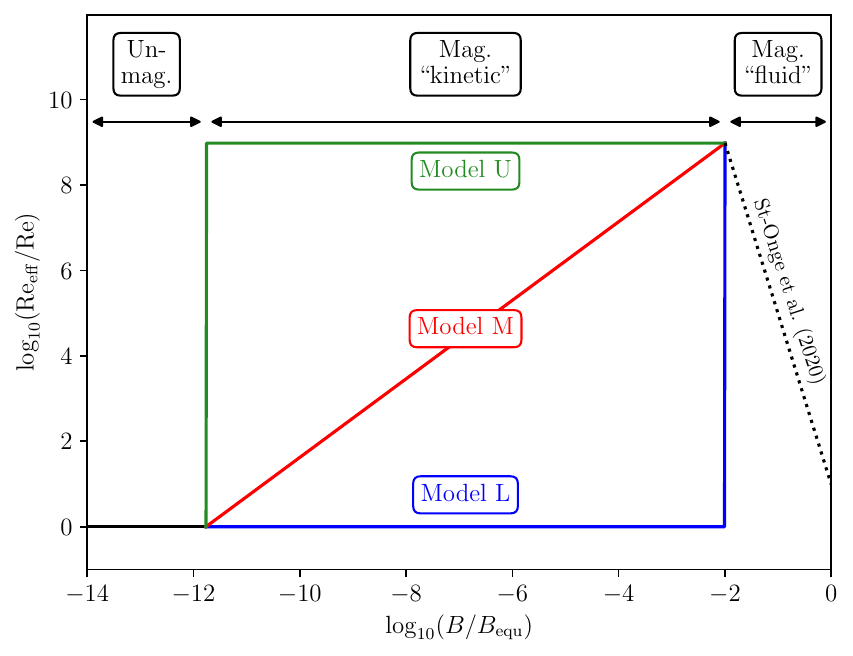}
    \caption{Representation of the different models for the effective Reynolds number for the following values: 
    $n_e = 10^{-3}~\mathrm{cm}^{-3},~\mathrm{Re} = 100,~T = 10^8~\mathrm{K},~v_{\mathrm{turb}}=300~\mathrm{km}/\mathrm{s},~v_{\mathrm{th}, i}= 1000~\mathrm{km}/\mathrm{s},~L_0 = 100~\mathrm{kpc}$ 
    and $B_{\mathrm{equ}} = 10^{-6}~\mathrm{G}$. 
    In the unmagnetized regime, the solid black line is the Reynolds 
    number that is not modified by the regulation of pressure anisotropies
    and is given in Eq.~\eqref{eq:classical_reynolds_number}.
    Between the limiting magnetic field strengths given in Eqns.~\eqref{eq:unmag_condition} and \eqref{eq:magfluid_limit}, 
    we construct the three different models, Model U, M, and L. 
    The black dotted line corresponds to results obtained from simulations 
    of weakly collisional plasmas presented in \citet{st-onge_kunz_squire_scheko_20202}.}
    \label{fig:effective_reynolds_number}
\end{figure}

To circumvent this issue, we have adopted the following strategy. 
We construct an effective Reynolds number for the three 
different magnetization states.
In the unmagnetized regime, the Reynolds number is given 
by its classical definition, i.e., Eq.~\eqref{eq:classical_reynolds_number}. 
In the magnetized fluid regime, we adopt the 
expression~\eqref{eq:effective_reynolds_number_magfluid}. 
However, due to the uncertain nature of the Reynolds number 
in the kinetic magnetized regime, we construct three different 
models (denoted L, M, and U, respectively), as shown in 
Fig.~\ref{fig:effective_reynolds_number}. This approach has been inspired by the qualitative illustration presented in Fig. 6.1 of \citet{st-onge_thesis}.
Our Model L considers a classical Reynolds number until reaching 
the value of the magnetized fluid regime. 
In Model M a linear interpolation between 
the unmagnetized and magnetized fluid regimes are assumed.
Finally, Model U considers a constant value throughout 
the kinetic magnetized regime, equal to the maximum 
Reynolds number in the fluid regime. 
By considering these different models, we aim to study
certain characteristics of dynamos operating in weakly collisional plasmas, 
such as the time required to reach equipartition with the 
turbulent kinetic energy. 
It should be noted that Model L only constitutes a lower limit on 
the effective Reynolds number in the situation where the effect of 
kinetic instabilities is to reduce the effective parallel viscosity 
by modifying the scattering of particles. 
In the case where the system is brought back to an equilibrium
state by reducing the stretching rate of the magnetic field 
(and hence increasing the effective viscosity), 
the value of the effective Reynolds number would decrease. 
Model U does not represent an upper limit on the Reynolds number either. 
We have established these models arbitrarily to study a wide 
range of values of the effective Reynolds number. 

Now, we have to establish the equation 
that governs the evolution and dynamics of the magnetic field.
In this paper, we assume that the magnetic field is 
amplified by the turbulent dynamo, which consists of the 
stretching, twisting, and folding of the magnetic field 
lines by turbulent motions \citep{kazantsev_1968}. 
If we consider the framework of Kolmogorov turbulence, 
the rate-of-strain tensor is dominated by motions 
at the viscous scale \citep{Schekochihin_2006_magfield_gclust}, 
which means $|\nabla \boldsymbol{u}| \approx (v_{\mathrm{turb}}/L_0)\mathrm{Re}^{1/2}$. 
On the other hand, in a plasma where a Braginskii-type viscosity is 
assumed, parallel motions to the magnetic field would be 
damped and only 
the parallel component of the rate-of-strain tensor $\hat{\boldsymbol{b}}\hat{\boldsymbol{b}} : \nabla \boldsymbol{u}$ is relevant. 
Therefore, using Eq.~(\ref{eq:delta_inst}) but replacing $\mathrm{Re}$ by $\mathrm{Re}_\mathrm{eff}$,
we obtain
\begin{equation}\label{eq:turbulent_dynamo}
\frac{1}{B}\frac{\mathrm{d} B}{\mathrm{d} t} \approx
\frac{v_{\mathrm{turb}}}{L_0}\mathrm{Re}_{\mathrm{eff}}^{1/2},
\end{equation}
where $\mathrm{Re}_{\mathrm{eff}}$ depends on the magnetization state and is given in Fig. \ref{fig:effective_reynolds_number}. 
We solve Eq.~\eqref{eq:turbulent_dynamo} for all magnetization states.
Once, the magnetic field approaches equipartition with the turbulent velocity field, nonlinear effects should set in and lead to saturation. 
However, modeling these nonlinear effects is beyond the scope of this paper, and we rather simply assume that the exponential growth ends when the magnetic field reaches the equipartition field strength $B_\mathrm{equ}$. 

All in all, we constructed $N_{\mathrm{tree}} = 10^3$ merger trees with 
$N_z = 300$ steps between $z=0$ and $z_{\mathrm{max}}$.
Once plasma parameters are calculated and averaged according to 
Eqs.~\eqref{eq:avg_dens} and \eqref{eq:avg_mz}, 
Eq.~\eqref{eq:turbulent_dynamo} is solved between all the redshift steps. Our choice of time resolution between two given redshift steps is explained in detail in Appendix \ref{appendix:time_res}.

% *********************************************************************
\section{Results}\label{sec:results}

\subsection{Basic plasma parameters from 1D profiles}

\begin{figure*}
    \centering
    \includegraphics[width = \textwidth]{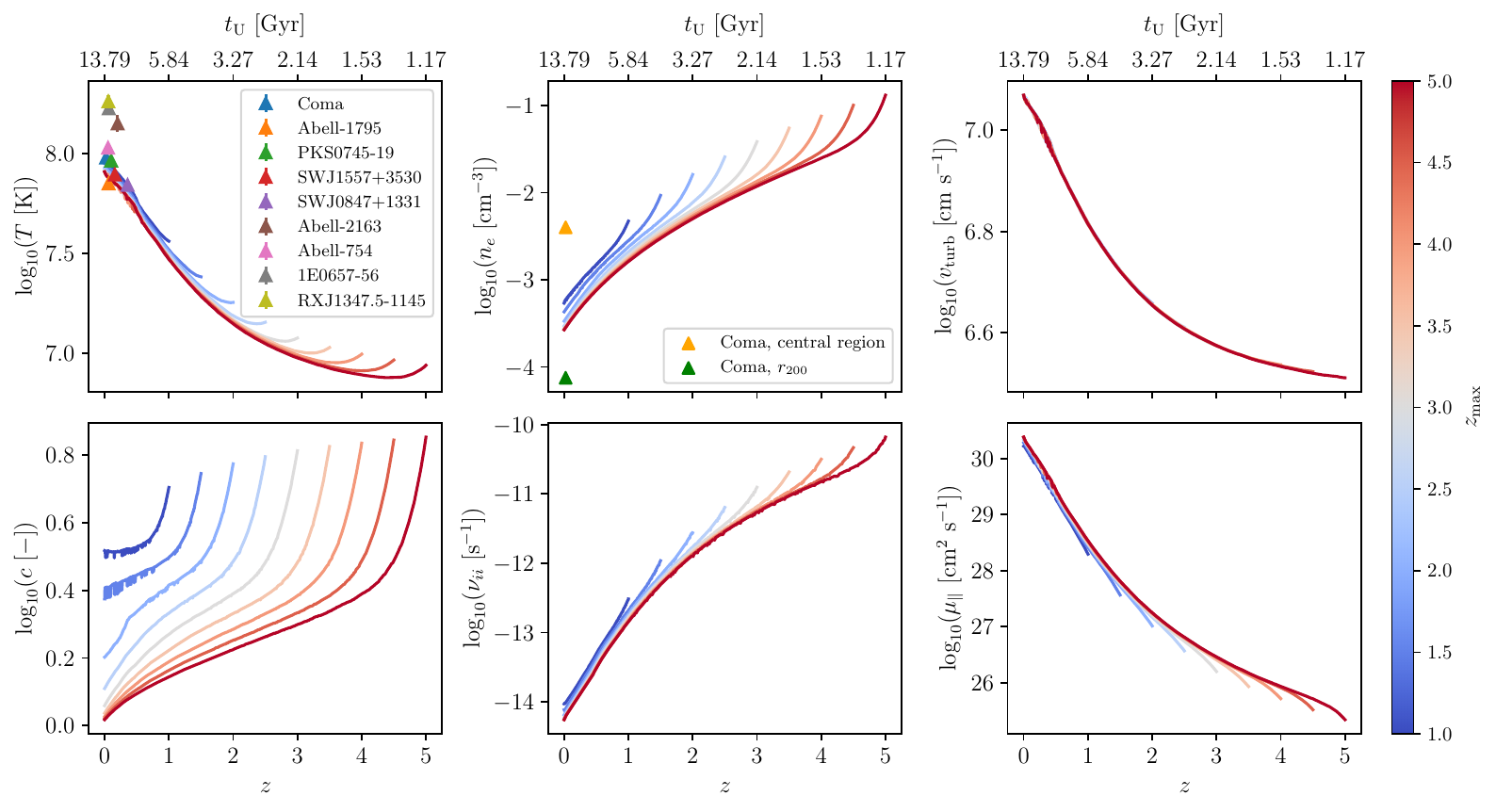}
    \caption{Evolution of various plasma quantities, as a function of redshift, and for different values of $z_{\mathrm{max}}$. Each curve represents the average value of the quantity over $N_{\mathrm{tree}} = 10^3$ merger trees (see Sec.~\ref{appendix:skew-normal}) for which a redshift-dependent, one-dimensional profile has been calculated according to Eqs.~\eqref{eq:avg_dens} 
    and \eqref{eq:avg_mz}.
    The age of the Universe $t_\mathrm{U}$ at each redshift is plotted on the second $x$-axis.
    In order to assess the validity of our model, we have added
    observed values for the temperature and 
    the thermal electron density
    (see the detailed list in the main text).}
    \label{fig:plasma_evolution_1}
\end{figure*}

Figure~\ref{fig:plasma_evolution_1} showcases the temporal evolution of various plasma parameters (Sec.~\ref{subsec:plasma_params}),
for different $z_{\mathrm{max}}$ values associated with merger trees generated using the Modified GALFORM algorithm.
Initially, we observe a similarity in the trends of the temperature and turbulent velocity curves.
The algorithm demonstrates a decline in these curves, by approximately 10 to 20\% of the total runtime, followed by a local minimum and a subsequent steady increase until reaching $z=0$.
Additionally, we notice that higher values of $z_{\mathrm{max}}$ lead to steeper local minima.
While the underlying cause of these curve shapes remains unclear, 
we can provide some explanations.
Firstly, it is important to note that the merger trees have been constructed with a uniform mass resolution of $M_{\mathrm{res}} = 10^{12}~M_{\odot}$, representing a thousandth of the final cluster's mass.
However, the GALFORM algorithm incorporates two stop conditions: reaching a predetermined value of $z_{\mathrm{max}}$ or the subhaloes reaching the resolution mass during the mass-splitting procedure, regardless of the redshift.
Consequently, a higher $z_{\mathrm{max}}$ corresponds to fewer subhaloes in the tree near this redshift.
Nonetheless, deviations from the mean values of temperature and turbulent velocity at $z=0$ remain minimal.
Furthermore, the concentration parameter $c$ is computed at each node of the tree by considering energy conservation, accounting for accreted matter.
When the merger tree extends to higher redshift values, the accretion of additional matter can substantially alter the final value of the concentration parameter, which could explain why $c$-values of the curves with $1 \lesssim z_{\mathrm{max}} \lesssim 2$ at $z=0$ are higher.
However, it is observed from Fig.~\ref{fig:plasma_evolution_1} that variations in $c$-values at $z=0$ do not seem to impact the electron density curves.
Remarkably, the $c$-value converges to approximately $c \simeq 1.2$ for $z \gtrsim 2.5$, being in relatively good agreement with the cMr-relation derived in \citet{biviano_2017_cmr_relation}.

In the preceding paragraph several concerns  
regarding the validity of our magnetic field model 
were raised,
as numerical artifacts arise for redshifts close to $z_{\mathrm{max}}$.
As previously stated, both the temperature and turbulent velocity curves 
exhibit a decreasing trend that precedes a global minimum and then 
displays a constant increase until $z=0$. 
However, it has been observed that this trend strongly relies on the initial conditions set by $z_{\mathrm{max}}$, resulting in a 
numerical effect in our model. In order to mitigate the potential impact of these artifacts on our conclusions, we have decided to impose constraints on the value of $z_{\mathrm{max}}$. Fig.~\ref{fig:plasma_evolution_1} clearly indicates that all curves converge towards a common value for redshift values $z\gtrsim 2$. Therefore, we have opted to exclude merger trees with $z_{\mathrm{max}} \leq 2$ for further analysis. Although selecting the merger tree with the highest $z_{\mathrm{max}}$ seems to be a plausible option, it is worth noting that the number of dark matter subhaloes decreases as $z_{\mathrm{max}}$ increases due to the way the GALFORM algorithm operates.
Therefore, from now on, we will only use the merger tree with $z_{\mathrm{max}} = 4$ to implement our magnetic field amplification model.
Additionally, we initiate the dynamo process at redshift values equal to or below $z_{\mathrm{start}} \leq 2$, ensuring that the influence of the mass resolution remains marginal.

Finally, in order to assess the robustness of our model, we have plotted on
Fig.~\ref{fig:plasma_evolution_1} 
observational values of various physical quantities for nine different clusters. 
It should be noted that this selection of clusters includes not only clusters 
in the same dynamical state; some are undergoing merging processes,
while others are in a relaxed state. 
For instance, \citet{miranda_2008_dynamics_RXJ1347} combined X-ray 
observations and strong lensing analysis of RX J1347.5-1145, and revealed 
a complex structure. 
They suggested a merger scenario between dark matter 
subclumps that would account for discrepancies with mass 
estimates from the virial theorem. 
Moreover, \citet{barrena_2002_dynamics_1E0657} provided evidence 
of a major collision between 1E0657-56 and a subcluster, 
and \citet{maurogordato_2008_dynamics_abell2163} provided 
optical observations suggesting that Abell 2163 had
undergone a recent ($\approx 0.5$ Gyr) collision.  
For the temperature, we have plotted results from X-ray observations 
for the Coma cluster, Abell 1795, PKS0745-19, SWJ1557+3530, and SWJ0847+1331 
from \citet{moretti_2011_temp_clusters} and references therein. 
We have also plotted the temperature from X-ray observations from \citet{wallbank_2022_xray_obs_temp} 
for Abell 2163, Abell 754, RX J1347.5-1145, and 1E 0657-56. 
Our model results in an average temperature at $z=0$ which is 
in good agreement with most clusters of the observational sample. 
We also present the thermal electron density 
values for the Coma cluster, both in its central region 
and its periphery, at $r = r_{200}$. 
These values were obtained from the X-ray morphology 
\citet{churazov_2021_coma} using the data from SRG/eROSITA. 
This enables us to assert that the typical plasma values 
expected at low redshift are overall in accordance with our model.

% /////////////////////////////////////////////////////////////////////////
\subsection{Magnetic field amplification}

In this section, we present the magnetic field amplification obtained from our model 
that is based on an effective Reynolds number. 
It is worth recalling that we have integrated our dynamo model into 
the merger tree with $z_{\mathrm{max}} = 4$, and initiated the amplification 
process at various redshifts between $z_{\mathrm{start}}=2$ and $z_{\mathrm{start}}=0.1$. 

\begin{figure*}
    \centering
    \includegraphics[width = \textwidth]{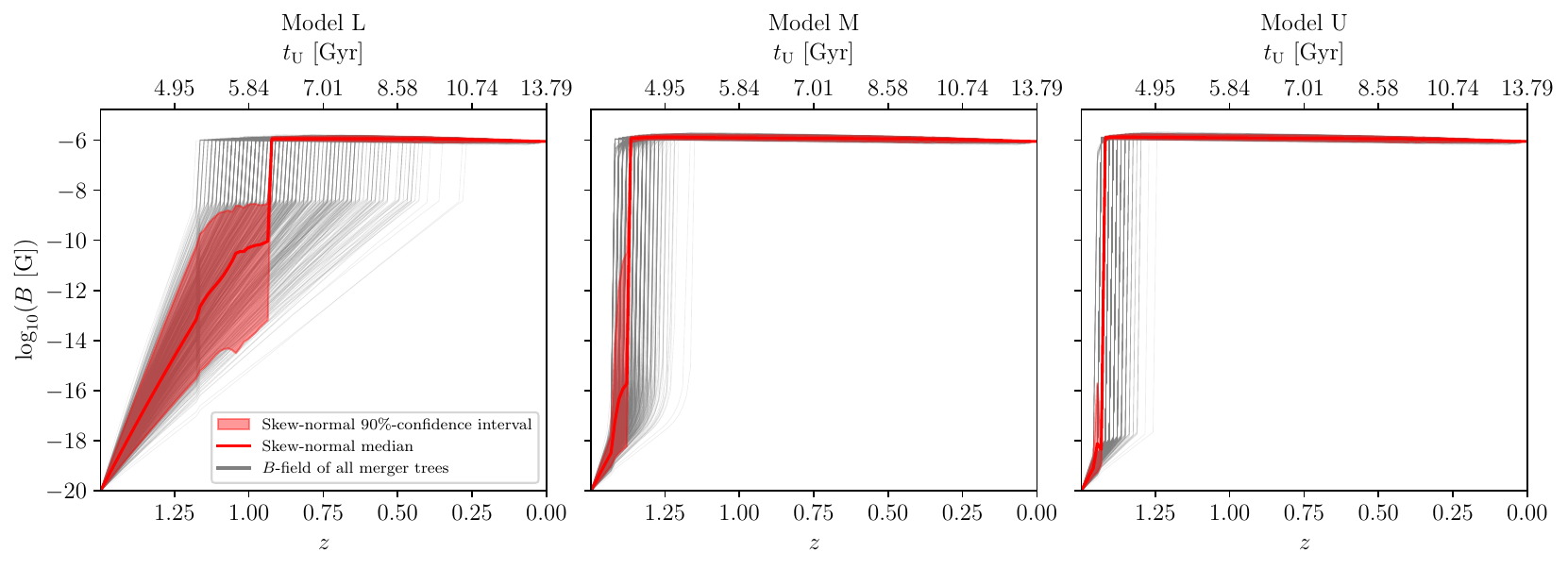}
    \caption{Evolution of the magnetic field, for $\alpha_0 = 20$ and $z_{\mathrm{start}} = 1.5$, and for the three models for the effective 
    Reynolds number. 
    The gray curves are the magnetic field evolution of all $N_{\mathrm{tree}} = 10^3$ merger trees. 
    % The area spanned by $N_{\mathrm{tree}} = 10^3$ merger trees is colored in gray. 
    The red curve indicates the average obtained by fitting all curves at each redshift value with a Skew-normal distribution (see Sec.~\ref{appendix:skew-normal}). The colored area shows the 90\%-confidence interval for such a distribution.
    The age of the Universe $t_\mathrm{U}$ at each redshift is plotted on the second $x$-axis.}
\label{fig:bfield_evolution_example_with_error}
\end{figure*}

\subsubsection{Example of magnetic field evolution}

Figure~\ref{fig:bfield_evolution_example_with_error} shows 
an example of the magnetic field evolution for the three 
effective Reynolds number models, detailed in 
Sec.~\ref{sec:model_magfields}, for $\alpha_0 = 20$ (see Eq.~\eqref{equ::inject_length} 
for the model of the turbulent forcing scale), 
$z_{\mathrm{start}} = 1.5$, and $B_0 = 10^{-20}~\mathrm{G}$. 
The gray curves represent the magnetic field evolution of all 
individual $N_{\mathrm{tree}} = 10^3$ merger trees.
In red, we show the Skew-normal average median with a 90\% confidence interval; 
see Appendix~\ref{appendix:skew-normal}.
Note that the estimated error during the exponential phase of 
Model L is larger than that of the other models. 
This can be partially explained by the fact that, in Model L, 
kinetic instabilities modify the Reynolds number 
only once the magnetic field amplitude reaches the order of 
$\mathrm{nG}$. 
Compared to Models M and U, in Model L, the Reynolds number remains at 
a value several orders of magnitude lower for a longer time, allowing the 
various curves to spread more until
the Reynolds number increases drastically, when $B$ exceeds the threshold given by Eq.~\eqref{eq:magfluid_limit}. 
In all three cases, Fig.~\ref{fig:bfield_evolution_example_with_error} demonstrates that once an amplitude of a few $\mathrm{nG}$ is achieved, the amplification of the magnetic field towards equipartition is nearly instantaneous.

\begin{figure*}[h]
    \centering
    \includegraphics[width = \textwidth]{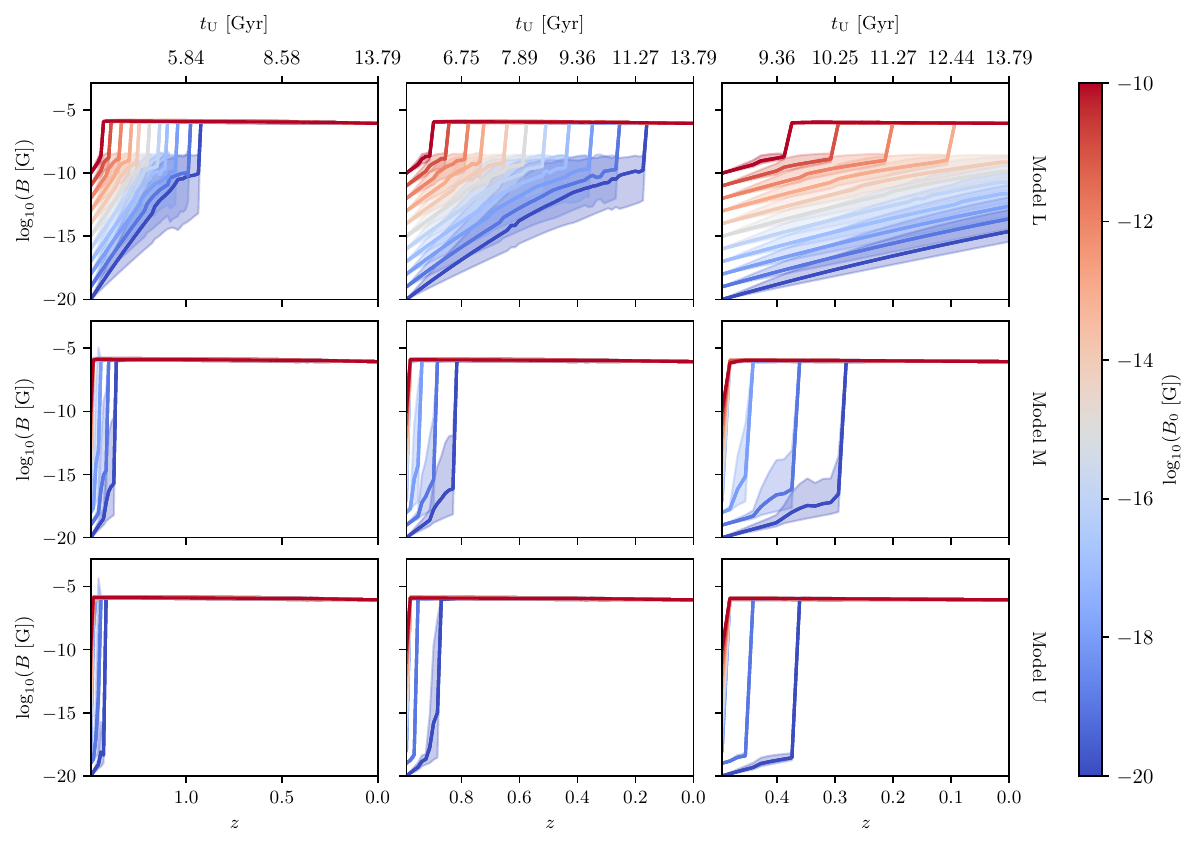}
    \caption{Evolution of the magnetic field amplitude over redshift for different models of effective Reynolds number.
    The curves and shaded areas correspond respectively to the average and the 90\%-confidence interval given by the averaging process described in Sec.~\ref{appendix:skew-normal}.
    The age of the Universe $t_\mathrm{U}$ at each redshift is plotted on the second $x$-axis.
    Each row corresponds to a different model, and each column corresponds to a different starting redshift of the dynamo. Model L (top row) has the lowest growth rate, while Models M and U (middle and bottom rows, respectively) show faster growth rates. The figure also shows that the dynamo was much faster in the past (higher redshift values). }
    \label{fig:bfield_evolution_example}
\end{figure*}

Figure \ref{fig:bfield_evolution_example} illustrates the 
time evolution 
of the magnetic field for the three distinct effective Reynolds number models, detailed in 
Sec.~\ref{sec:model_magfields}, 
for a specific value of $\alpha_0 = 20$, starting at redshifts of $z_{\mathrm{start}}=0.5$, $1$, and $1.5$.
Figure~\ref{fig:bfield_evolution_example} illustrates the same evolution of the magnetic field as on Fig.~\ref{fig:bfield_evolution_example_with_error}, but this time for $z_{\mathrm{start}} = 0.5,1,1.5$, and for initial seed field values ranging from $\log_{10}(B~[\mathrm{G}]) = -20$ to $-10$.
Firstly, it is evident that for models M and U, the curves coincide for $B_0 \gtrsim 10^{-16}~\mathrm{G}$
and, in particular, the time required to reach equipartition with the turbulent velocity field is independent of $B_0$. 
Therefore, we choose to omit the error bars in our subsequent analyses described below.
It also appears that the growth rate of the magnetic field becomes faster as the redshift increases.
This effect can only be justified by the evolution of various plasma parameters incorporated in 
Eq.~\eqref{eq:turbulent_dynamo} 
for the growth rate.

% /////////////////////////////////////////////////////////////////////////
\subsubsection{Time to equipartition}
\begin{figure*}
    \centering
    \includegraphics[width = \textwidth]{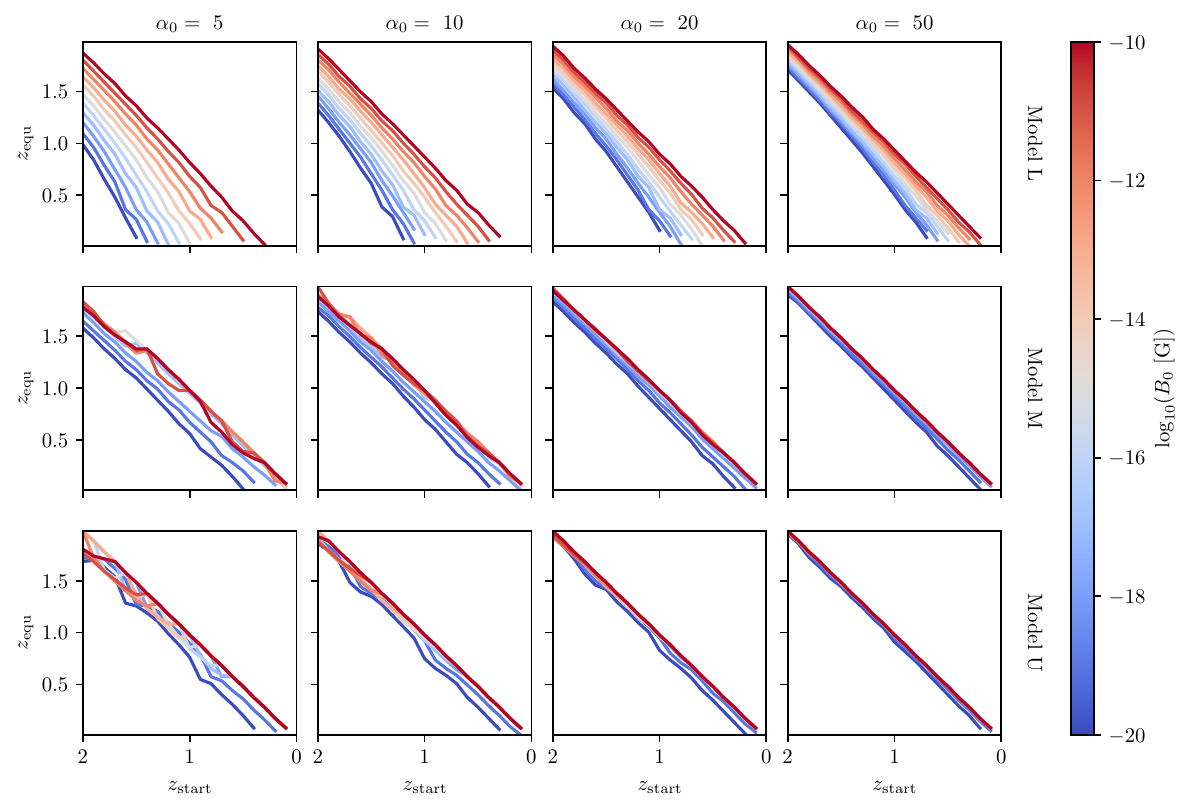}
    \caption{Redshift $z_{\mathrm{equ}}$ at which the magnetic field reaches at least 90\% of equipartition with the turbulent velocity field, for different values of the initial magnetic seed field.
    Each row corresponds to the different models of the effective Reynolds number in the magnetized kinetic regime (Sec. \ref{sec:model_magfields}). Each column corresponds to a different turbulent injection scale.}
    \label{fig:time_to_equipartition}
\end{figure*}

\begin{figure}
    \centering
\includegraphics[width = \columnwidth]{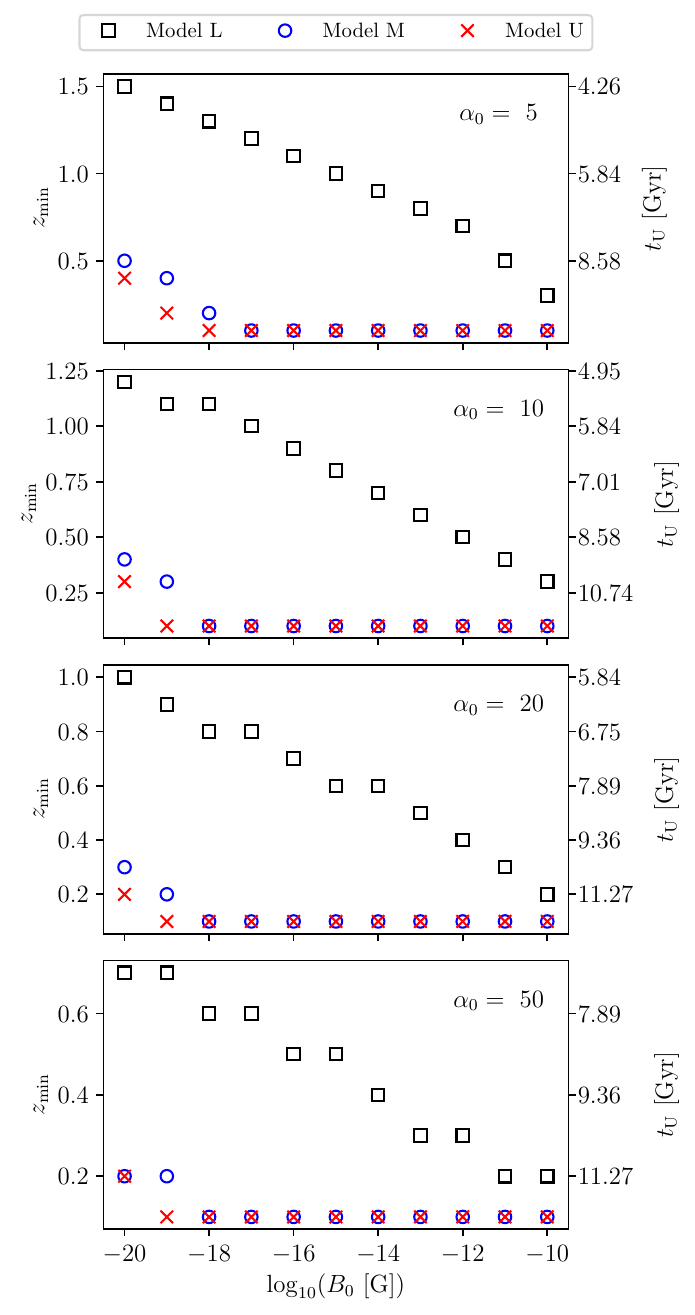}
\caption{Minimum redshift $z_{\mathrm{min}}$ from which a dynamo has to start to take the magnetic field $B$ up to equipartition with the turbulent velocity field at $z=0$, for different values of the initial magnetic seed field, and for each model for the effective Reynolds number. Each panel corresponds to a specific value of the injection scale defined by Eq.~\eqref{equ::inject_length}.}
\label{fig:limit_redshift_to_equipartition}
\end{figure}

Figure~\ref{fig:time_to_equipartition} shows the values of redshift $z_{\mathrm{equ}}$ at which the magnetic energy density reaches at least 90\% of the kinetic energy density of turbulence, for the three models of $\mathrm{Re}_{\mathrm{eff}}$, and for different values of $\alpha_0$, $B_0$ and $z_{\mathrm{start}}$.
Figure~\ref{fig:limit_redshift_to_equipartition} 
shows the minimum redshift $z_{\mathrm{min}}$ at which the dynamo has to 
start such that equipartition with the turbulent velocity field can be reached, for the three effective Reynolds number models and for various values of $\alpha_0$. 
Firstly, it can be observed that equipartition can be reached in all the different configurations that we have implemented.
Moreover, it is clear that Models M and U are indistinguishable for 
$\log_{10}(B_0~[\mathrm{G}]) \gtrsim -17$.
Furthermore, below these values, the differences between their respective curves are not significant.

The important conclusions that can be drawn from these results are as follows.
Firstly, assuming that the modification of effective ion-ion 
collisionality is the sole physical quantity modified by
kinetic instabilities, it can reasonably be accepted that 
Model L represents a lower limit of $z_{\mathrm{min}}$.
Additionally, even though Model U does not represent an
upper limit to the effective Reynolds number, it can be assumed that any faster model would not result in any changes to the values presented in Fig.~\ref{fig:limit_redshift_to_equipartition}.
In summary, any effective model for the 
Reynolds number lying between the curves associated with 
Models M and U would lead to an explosive dynamo so fast 
that their evolution could not be distinguished with our 
resolution in redshift.
On the other hand, considering a model between the curves L and M would indeed lead to changes in the values of $z_{\mathrm{min}}$ compared to those presented in Fig.~\ref{fig:limit_redshift_to_equipartition}.
Finally, we are limited in the conclusions we can draw due to our simple limitation in modeling the dynamics of the turbulent field, particularly its injection scale.
If $\alpha_0$ were to evolve, especially through certain mechanisms of turbulent decay, our results would obviously be modified.
We leave a more comprehensive modeling of the turbulent velocity field for future work.

% *********************************************************************
\section{Discussion}\label{sec:discussions}
Using merger trees has enabled us to study the 
evolution of certain plasma quantities, 
including the magnetic field, without relying 
on computationally expensive simulations.
However, such an approach presents a number 
of uncertainties. 
Firstly, we have only used the Modified GALFORM algorithm, 
as it has been calibrated to the mass function 
of the Millennium simulations.
We do not know (yet) if our conclusions would change 
when considering another merger tree algorithm.

When applying the merger tree algorithm, we have to 
pay special attention to numerical artifacts. 
In particular, we have constrained the value of $z_{\mathrm{max}}$ 
(redshift at which the merging process starts) to $z_{\mathrm{max}} = 4$, 
so that our results are not influenced by numerical effects related 
to mass resolution (see Fig.~\ref{fig:plasma_evolution_1}). 
Furthermore, we decided to start the dynamo process from 
$z_{\mathrm{start}}\simeq 2$ or below. 
Indeed, below $z_{\mathrm{start}} = 2$, all curves with 
$z_{\mathrm{max}} \geq 4$ are approximately identical.
Therefore, our ability to study 
dynamo processes at $z \gtrsim 2$ depends on the 
model for the concentration parameter considered.

Also, the effective Reynolds number has to be chosen carefully,
mainly because Models L and 
U do not necessarily constitute a lower and upper 
bound per se.
This may be susceptible to introducing biases in the 
minimum redshift, $z_\mathrm{min}$, at which the magnetic 
field amplification needs to start
in order to reach equipartition at $z=0$.
However, our results have shown that Models M and 
U produce nearly identical results.
This is not surprising, because even if Models M and U 
are not the same, they both differ from the classical 
Reynolds number by (many) orders of magnitude.

Furthermore, confining the evolution of the turbulent injection 
scale appears to be one of the few 
necessary conditions to constrain the evolution of 
the magnetic field. 
It should be noted that other sources of turbulence are 
likely to exist during certain phases of the formation of a cluster,
potentially with different forcing length scales. 
Other sources include, for example, subcluster and galactic wakes 
\citep[e.g.][]{subramanian_2006} and jets from active galactic 
nuclei \citep[e.g.][]{fujita_2020_agn}.

Finally, for the scenario presented in Fig.~\ref{fig:bfield_evolution_example_with_error}
for instance, the magnetic field strength increases by 14 orders of magnitude,
while the temperature increases by one order of magnitude. 
In this specific example, the $\beta$ parameter changes from $\beta \approx 10^{28}$
at $z=1.5$ to $\beta \approx 10^2$ at $z=0$. 
Therefore we do not expect direct feedback on cluster evolution from the magnetic field through magnetic pressure. 
However, even at times where $\beta \gg 1$, the strength and direction of the viscosity are affected by the instabilities. 
This implies that indirect feedback from the magnetic field on the flow can occur,
leading to effects that are not captured by classical hydrodynamic simulations 
(see also the discussion in \citet{st-onge_2018_hybrid}).

% *********************************************************************
\section{Conclusions}\label{sec:conclusions}
In this work, we used the Modified GALFORM code \citep{parkinson_2008_mod_galform} to investigate the amplification of magnetic fields in the intracluster medium (ICM) during the formation of galaxy clusters by successive mergers of dark matter halos. We focused on clusters with a typical mass of $M = 10^{15}~\mathrm{M}_{\odot}$. 
To this end, we implemented one-dimensional profiles for the distribution of dark matter, baryonic gas, temperature, and turbulent velocity, assuming spherical symmetry, and following the methodology proposed by \citet{dvorkin_Rephaeli_2011} and \citet{Johnson_2021_random_walk_c}. 
For a given merger tree realization, the profiles of all subhalos at a given redshift were averaged in order to calculate relevant plasma parameters as a function of redshift. This process has been performed for $N_{\mathrm{tree}} = 10^3$ merger trees to establish statistical properties.
We compared the temperature at $z=0$ obtained from our model with observed values from a sample of selected clusters presented in \citet{moretti_2011_temp_clusters} and \citet{wallbank_2022_xray_obs_temp}.
% We also compared other quantities like electron density, turbulent velocity, and ion-ion collision frequency to typical values deduced from Hydra A observations listed in \citet{Schekochihin_2006_magfield_gclust}. 
We also compared the electron density values of our model with SRG/eROSITA
observations of the Coma cluster from \citet{churazov_2021_coma}.
Overall, we found our model to be in good agreement with the observed values 
from a sample of low-redshift clusters.  

We then constructed a semi-analytical model for the 
amplification of the magnetic field that incorporates the
effects of pressure-anisotropy-driven kinetic instabilities 
(firehose and mirror) for different magnetization regimes \citep{melville_2016,Rincon_2016_colless_dynamo, st-onge_2018_hybrid, st-onge_kunz_squire_scheko_20202}. 
Next, we estimated the growth rate of the magnetic field for different values of 
the strength of the magnetic seed field, and 
forcing length scales of turbulence.
This allowed us to determine the redshift at which equipartition is reached in different scenarios. In practice, we define equipartition as the time when the magnetic energy density reaches at least 90\% of the turbulent kinetic energy density.
Specifically, for an injection scale of $L_0 = r_{\mathrm{vir}}/20$, where $r_{\mathrm{vir}}$ is the virial radius of the dark matter halo, the magnetic field evolves in the following way.
For our slowest dynamo (effect Reynolds number Model L), the dynamo 
amplification for seed field values of $\log_{10}(B_0~[\mathrm{G}]) = -20$ and $-10$ 
has to start at least from, respectively, $z_{\mathrm{start}} \simeq 1$ and $0.2$ to reach 
equipartition until the present day. 
For the other faster scenarios (Models M and U), those values are respectively 
reduced to $z_{\mathrm{start}} \simeq 0.3$ and $\simeq 0.1$.
Although our Model U cannot be considered an upper-limit 
for the effective Reynolds number, 
our Model L can constitute a lower limit, 
provided that the increase of ion-ion collisionality is the sole 
effect created by the kinetic instabilities to bring the system 
back to marginal stability.

Overall, our study demonstrates that merger trees can be a 
valuable tool for constraining the evolution of magnetic fields 
over cosmological times, particularly in galaxy clusters. 
In particular, they allow us to probe kinetic plasma effects in the
history of the ICM which are inaccessible with state-of-the-art
cosmological simulations.
Although our approach is not as accurate as fully kinetic plasma simulations, 
it could potentially constitute a computationally efficient alternative for 
constraining the dependence of the effective Reynolds number on the magnetic 
field, in the magnetic ``kinetic'' regime, when combined with future and 
more resolved radio observations.

% \bibliographystyle{aa}
% \bibliography{biblio.bib}

\begin{thebibliography}{89}
\expandafter\ifx\csname natexlab\endcsname\relax\def\natexlab#1{#1}\fi

\bibitem[{{Adshead} {et~al.}(2016){Adshead}, {Giblin}, {Scully}, \&
  {Sfakianakis}}]{adshead_2016_magnetogenesis_axion}
{Adshead}, P., {Giblin}, John~T., J., {Scully}, T.~R., \& {Sfakianakis}, E.~I.
  2016, \jcap, 2016, 039

\bibitem[{{Barrena} {et~al.}(2002){Barrena}, {Biviano}, {Ramella}, {Falco}, \&
  {Seitz}}]{barrena_2002_dynamics_1E0657}
{Barrena}, R., {Biviano}, A., {Ramella}, M., {Falco}, E.~E., \& {Seitz}, S.
  2002, \aap, 386, 816

\bibitem[{{Beck} \& {Wielebinski}(2013)}]{bfields_galaxies_beck_2013}
{Beck}, R. \& {Wielebinski}, R. 2013, in Planets, Stars and Stellar Systems.
  Volume 5: Galactic Structure and Stellar Populations, ed. T.~D. {Oswalt} \&
  G.~{Gilmore}, Vol.~5, 641

\bibitem[{{Beresnyak}(2012)}]{beresnyak_2012_universal_ssd}
{Beresnyak}, A. 2012, \prl, 108, 035002

\bibitem[{{Bernet} {et~al.}(2008){Bernet}, {Miniati}, {Lilly}, {Kronberg}, \&
  {Dessauges-Zavadsky}}]{bernet_2008_strong_bfields_gal}
{Bernet}, M.~L., {Miniati}, F., {Lilly}, S.~J., {Kronberg}, P.~P., \&
  {Dessauges-Zavadsky}, M. 2008, \nat, 454, 302

\bibitem[{{Biermann}(1950)}]{biermann_1950}
{Biermann}, L. 1950, Zeitschrift Naturforschung Teil A, 5, 65

\bibitem[{{Biviano} {et~al.}(2017){Biviano}, {Moretti}, {Paccagnella},
  {Poggianti}, {Bettoni}, {Gullieuszik}, {Vulcani}, {Fasano}, {D'Onofrio},
  {Fritz}, \& {Cava}}]{biviano_2017_cmr_relation}
{Biviano}, A., {Moretti}, A., {Paccagnella}, A., {et~al.} 2017, \aap, 607, A81

\bibitem[{{Bonafede} {et~al.}(2010){Bonafede}, {Feretti}, {Murgia}, {Govoni},
  {Giovannini}, {Dallacasa}, {Dolag}, \&
  {Taylor}}]{bionafede_2010_rot_measure_coma}
{Bonafede}, A., {Feretti}, L., {Murgia}, M., {et~al.} 2010, \aap, 513, A30

\bibitem[{{Braginskii}(1965)}]{braginskii_1965}
{Braginskii}, S.~I. 1965, Reviews of Plasma Physics, 1, 205

\bibitem[{{Brandenburg} {et~al.}(2023){Brandenburg}, {Rogachevskii}, \&
  {Schober}}]{brandenburg2023_diss_ssd}
{Brandenburg}, A., {Rogachevskii}, I., \& {Schober}, J. 2023, \mnras, 518, 6367

\bibitem[{Brandenburg {et~al.}(2017)Brandenburg, Schober, Rogachevskii,
  Kahniashvili, Boyarsky, Fröhlich, Ruchayskiy, \&
  Kleeorin}]{Brandenburg_2017_chiral_effect_early_universe}
Brandenburg, A., Schober, J., Rogachevskii, I., {et~al.} 2017, The
  Astrophysical Journal Letters, 845, L21

\bibitem[{{Chandrasekhar} {et~al.}(1958{\natexlab{a}}){Chandrasekhar},
  {Kaufman}, \& {Watson}}]{rosenbluth_1956_pinch}
{Chandrasekhar}, S., {Kaufman}, A.~N., \& {Watson}, K.~M. 1958{\natexlab{a}},
  Proceedings of the Royal Society of London Series A, 245, 435

\bibitem[{{Chandrasekhar} {et~al.}(1958{\natexlab{b}}){Chandrasekhar},
  {Kaufman}, \& {Watson}}]{CKW_1958_pinch}
{Chandrasekhar}, S., {Kaufman}, A.~N., \& {Watson}, K.~M. 1958{\natexlab{b}},
  Proceedings of the Royal Society of London Series A, 245, 435

\bibitem[{{Chew} {et~al.}(1956){Chew}, {Goldberger}, \&
  {Low}}]{chew_goldberger_low_1956_closure}
{Chew}, G.~F., {Goldberger}, M.~L., \& {Low}, F.~E. 1956, Proceedings of the
  Royal Society of London Series A, 236, 112

\bibitem[{{Churazov} {et~al.}(2021){Churazov}, {Khabibullin}, {Lyskova},
  {Sunyaev}, \& {Bykov}}]{churazov_2021_coma}
{Churazov}, E., {Khabibullin}, I., {Lyskova}, N., {Sunyaev}, R., \& {Bykov},
  A.~M. 2021, \aap, 651, A41

\bibitem[{{Cole} {et~al.}(2000){Cole}, {Lacey}, {Baugh}, \&
  {Frenk}}]{cole2000_galform}
{Cole}, S., {Lacey}, C.~G., {Baugh}, C.~M., \& {Frenk}, C.~S. 2000, \mnras,
  319, 168

\bibitem[{{Di Gennaro} {et~al.}(2021){Di Gennaro}, {van Weeren}, {Brunetti},
  {Cassano}, {Br{\"u}ggen}, {Hoeft}, {Shimwell}, {R{\"o}ttgering}, {Bonafede},
  {Botteon}, {Cuciti}, {Dallacasa}, {de Gasperin},
  {Dom{\'\i}nguez-Fern{\'a}ndez}, {En{\ss}lin}, {Gastaldello}, {Mandal},
  {Rossetti}, \& {Simionescu}}]{di_gennaro_2020}
{Di Gennaro}, G., {van Weeren}, R.~J., {Brunetti}, G., {et~al.} 2021, Nature
  Astronomy, 5, 268

\bibitem[{Domínguez-Fernández {et~al.}(2019)Domínguez-Fernández, Vazza,
  Brüggen, \& Brunetti}]{Domínguez-Fernandez_2019_cluster}
Domínguez-Fernández, P., Vazza, F., Brüggen, M., \& Brunetti, G. 2019,
  Monthly Notices of the Royal Astronomical Society, 486, 623

\bibitem[{Dvorkin \& Rephaeli(2011)}]{dvorkin_Rephaeli_2011}
Dvorkin, I. \& Rephaeli, Y. 2011, \mnras, 412, 665

\bibitem[{{Ellis} {et~al.}(2019){Ellis}, {Fairbairn}, {Lewicki}, {Vaskonen}, \&
  {Wickens}}]{ellis_2019_phase_trans_bfields}
{Ellis}, J., {Fairbairn}, M., {Lewicki}, M., {Vaskonen}, V., \& {Wickens}, A.
  2019, \jcap, 2019, 019

\bibitem[{Fitzpatrick(2014)}]{fitzpatrick2014plasma}
Fitzpatrick, R. 2014, Plasma Physics: An Introduction (Taylor \& Francis)

\bibitem[{{Fujita} {et~al.}(2015){Fujita}, {Namba}, {Tada}, {Takeda}, \&
  {Tashiro}}]{fujita_2015_magnetogenesis_axion}
{Fujita}, T., {Namba}, R., {Tada}, Y., {Takeda}, N., \& {Tashiro}, H. 2015,
  \jcap, 2015, 054

\bibitem[{{Fujita} {et~al.}(2020){Fujita}, {Cen}, \&
  {Zhuravleva}}]{fujita_2020_agn}
{Fujita}, Y., {Cen}, R., \& {Zhuravleva}, I. 2020, \mnras, 494, 5507

\bibitem[{{Gary}(1992)}]{gary_1992}
{Gary}, S.~P. 1992, \jgr, 97, 8519

\bibitem[{{Gnedin} {et~al.}(2000){Gnedin}, {Ferrara}, \&
  {Zweibel}}]{gnedin_2000_reionization}
{Gnedin}, N.~Y., {Ferrara}, A., \& {Zweibel}, E.~G. 2000, \apj, 539, 505

\bibitem[{{G{\'o}mez} {et~al.}(2022){G{\'o}mez}, {Padilla}, {Helly}, {Lacey},
  {Baugh}, \& {Lagos}}]{gomez_2022_mt_galaxies}
{G{\'o}mez}, J.~S., {Padilla}, N.~D., {Helly}, J.~C., {et~al.} 2022, \mnras,
  510, 5500

\bibitem[{Helander {et~al.}(2016)Helander, Strumik, \&
  Schekochihin}]{helander_strumik_schekochihin_2016}
Helander, P., Strumik, M., \& Schekochihin, A.~A. 2016, Journal of Plasma
  Physics, 82, 905820601

\bibitem[{{Hellinger}(2007)}]{hellinge_2007}
{Hellinger}, P. 2007, Physics of Plasmas, 14, 082105

\bibitem[{{Hubrig} {et~al.}(2011){Hubrig}, {Sch{\"o}ller}, {Kharchenko},
  {Langer}, {de Wit}, {Ilyin}, {Kholtygin}, {Piskunov}, {Przybilla}, \& {Magori
  Collaboration}}]{hubrig_2011_bfields_massive_stars}
{Hubrig}, S., {Sch{\"o}ller}, M., {Kharchenko}, N.~V., {et~al.} 2011, \aap,
  528, A151

\bibitem[{{Jedamzik} \& {Saveliev}(2019)}]{jedamzik_2019_CMB_limit}
{Jedamzik}, K. \& {Saveliev}, A. 2019, \prl, 123, 021301

\bibitem[{Johnson {et~al.}(2021)Johnson, Benson, \&
  Grin}]{Johnson_2021_random_walk_c}
Johnson, T., Benson, A.~J., \& Grin, D. 2021, The Astrophysical Journal, 908,
  33

\bibitem[{{Joyce} \& {Shaposhnikov}(1997)}]{joyce_1997_primordial_bfields}
{Joyce}, M. \& {Shaposhnikov}, M. 1997, Phys.~Rev.~Lett., 79, 1193

\bibitem[{{Kazantsev}(1968)}]{kazantsev_1968}
{Kazantsev}, A.~P. 1968, Soviet Journal of Experimental and Theoretical
  Physics, 26, 1031

\bibitem[{Kravtsov \& Borgani(2012)}]{ann_ewv_galaxy_clusters_formation}
Kravtsov, A.~V. \& Borgani, S. 2012, Annual Review of Astronomy and
  Astrophysics, 50, 353

\bibitem[{{Kulsrud} {et~al.}(1997{\natexlab{a}}){Kulsrud}, {Cowley},
  {Gruzinov}, \& {Sudan}}]{Kulsrud_1997_cosmic_dynamos}
{Kulsrud}, R., {Cowley}, S.~C., {Gruzinov}, A.~V., \& {Sudan}, R.~N.
  1997{\natexlab{a}}, \physrep, 283, 213

\bibitem[{Kulsrud(1983)}]{kulsrud_1983_MHD_description}
Kulsrud, R.~M. 1983, in Handbook of plasma physics. Vol. 1: Basic plasma
  physics I., ed. A.~A. Galeev \& R.~N. Sudan, 115--144

\bibitem[{{Kulsrud} {et~al.}(1997{\natexlab{b}}){Kulsrud}, {Cen}, {Ostriker},
  \& {Ryu}}]{biermann_kulsrud_1997}
{Kulsrud}, R.~M., {Cen}, R., {Ostriker}, J.~P., \& {Ryu}, D.
  1997{\natexlab{b}}, \apj, 480, 481

\bibitem[{{Kunz} {et~al.}(2014){Kunz}, {Schekochihin}, \&
  {Stone}}]{kunz_scheko_stone_2014}
{Kunz}, M.~W., {Schekochihin}, A.~A., \& {Stone}, J.~M. 2014, \prl, 112, 205003

\bibitem[{{Marinacci} {et~al.}(2018){Marinacci}, {Vogelsberger}, {Pakmor},
  {Torrey}, {Springel}, {Hernquist}, {Nelson}, {Weinberger}, {Pillepich},
  {Naiman}, \& {Genel}}]{marinacci_illustris_2018}
{Marinacci}, F., {Vogelsberger}, M., {Pakmor}, R., {et~al.} 2018, \mnras, 480,
  5113

\bibitem[{{Maurogordato} {et~al.}(2008){Maurogordato}, {Cappi}, {Ferrari},
  {Benoist}, {Mars}, {Soucail}, {Arnaud}, {Pratt}, {Bourdin}, \&
  {Sauvageot}}]{maurogordato_2008_dynamics_abell2163}
{Maurogordato}, S., {Cappi}, A., {Ferrari}, C., {et~al.} 2008, \aap, 481, 593

\bibitem[{{McCarthy} {et~al.}(2007){McCarthy}, {Bower}, \&
  {Balogh}}]{mccarthy_2007_baryon_frac_clusters}
{McCarthy}, I.~G., {Bower}, R.~G., \& {Balogh}, M.~L. 2007, \mnras, 377, 1457

\bibitem[{{Melville} {et~al.}(2016){Melville}, {Schekochihin}, \&
  {Kunz}}]{melville_2016}
{Melville}, S., {Schekochihin}, A.~A., \& {Kunz}, M.~W. 2016, \mnras, 459, 2701

\bibitem[{{Miniati} \& {Beresnyak}(2015)}]{Miniati_2015_ICM_dynamo}
{Miniati}, F. \& {Beresnyak}, A. 2015, \nat, 523, 59

\bibitem[{{Miranda} {et~al.}(2008){Miranda}, {Sereno}, {de Filippis}, \&
  {Paolillo}}]{miranda_2008_dynamics_RXJ1347}
{Miranda}, M., {Sereno}, M., {de Filippis}, E., \& {Paolillo}, M. 2008, \mnras,
  385, 511

\bibitem[{{Mo} {et~al.}(2010){Mo}, {van den Bosch}, \&
  {White}}]{mo_galaxy_formation_notebook}
{Mo}, H., {van den Bosch}, F.~C., \& {White}, S. 2010, {Galaxy Formation and
  Evolution}

\bibitem[{{Mogavero} \& {Schekochihin}(2014)}]{mogavero_scheko_2014}
{Mogavero}, F. \& {Schekochihin}, A.~A. 2014, \mnras, 440, 3226

\bibitem[{{Moretti} {et~al.}(2011){Moretti}, {Gastaldello}, {Ettori}, \&
  {Molendi}}]{moretti_2011_temp_clusters}
{Moretti}, A., {Gastaldello}, F., {Ettori}, S., \& {Molendi}, S. 2011, \aap,
  528, A102

\bibitem[{{Naoz} \& {Narayan}(2013)}]{naoz_narayan_2013}
{Naoz}, S. \& {Narayan}, R. 2013, \prl, 111, 051303

\bibitem[{{Navarro} {et~al.}(1996){Navarro}, {Frenk}, \&
  {White}}]{navarro_frenk_white_1996}
{Navarro}, J.~F., {Frenk}, C.~S., \& {White}, S. D.~M. 1996, \apj, 462, 563

\bibitem[{{Neto} {et~al.}(2007){Neto}, {Gao}, {Bett}, {Cole}, {Navarro},
  {Frenk}, {White}, {Springel}, \& {Jenkins}}]{neto_2007_cold_dm_statistics}
{Neto}, A.~F., {Gao}, L., {Bett}, P., {et~al.} 2007, \mnras, 381, 1450

\bibitem[{{Ostriker} {et~al.}(2005){Ostriker}, {Bode}, \&
  {Babul}}]{ostriker_2005_T_and_rho}
{Ostriker}, J.~P., {Bode}, P., \& {Babul}, A. 2005, \apj, 634, 964

\bibitem[{{Parker}(1958)}]{parker_1958_aniso_gas}
{Parker}, E.~N. 1958, Physical Review, 109, 1874

\bibitem[{{Parkinson} {et~al.}(2008){Parkinson}, {Cole}, \&
  {Helly}}]{parkinson_2008_mod_galform}
{Parkinson}, H., {Cole}, S., \& {Helly}, J. 2008, \mnras, 383, 557

\bibitem[{{Planck Collaboration} {et~al.}(2016){Planck Collaboration}, {Ade},
  {Aghanim}, {Arnaud}, {Arroja}, {Ashdown}, {Aumont}, {Baccigalupi},
  {Ballardini}, {Banday}, {Barreiro}, {Bartolo}, {Battaner}, {Benabed},
  {Beno{\^\i}t}, {Benoit-L{\'e}vy}, {Bernard}, {Bersanelli}, {Bielewicz},
  {Bock}, {Bonaldi}, {Bonavera}, {Bond}, {Borrill}, {Bouchet}, {Bucher},
  {Burigana}, {Butler}, {Calabrese}, {Cardoso}, {Catalano}, {Chamballu},
  {Chiang}, {Chluba}, {Christensen}, {Church}, {Clements}, {Colombi},
  {Colombo}, {Combet}, {Couchot}, {Coulais}, {Crill}, {Curto}, {Cuttaia},
  {Danese}, {Davies}, {Davis}, {de Bernardis}, {de Rosa}, {de Zotti},
  {Delabrouille}, {D{\'e}sert}, {Diego}, {Dolag}, {Dole}, {Donzelli},
  {Dor{\'e}}, {Douspis}, {Ducout}, {Dupac}, {Efstathiou}, {Elsner},
  {En{\ss}lin}, {Eriksen}, {Fergusson}, {Finelli}, {Florido}, {Forni},
  {Frailis}, {Fraisse}, {Franceschi}, {Frejsel}, {Galeotta}, {Galli}, {Ganga},
  {Giard}, {Giraud-H{\'e}raud}, {Gjerl{\o}w}, {Gonz{\'a}lez-Nuevo},
  {G{\'o}rski}, {Gratton}, {Gregorio}, {Gruppuso}, {Gudmundsson}, {Hansen},
  {Hanson}, {Harrison}, {Helou}, {Henrot-Versill{\'e}},
  {Hern{\'a}ndez-Monteagudo}, {Herranz}, {Hildebrandt}, {Hivon}, {Hobson},
  {Holmes}, {Hornstrup}, {Hovest}, {Huffenberger}, {Hurier}, {Jaffe}, {Jaffe},
  {Jones}, {Juvela}, {Keih{\"a}nen}, {Keskitalo}, {Kim}, {Kisner}, {Knoche},
  {Kunz}, {Kurki-Suonio}, {Lagache}, {L{\"a}hteenm{\"a}ki}, {Lamarre},
  {Lasenby}, {Lattanzi}, {Lawrence}, {Leahy}, {Leonardi}, {Lesgourgues},
  {Levrier}, {Liguori}, {Lilje}, {Linden-V{\o}rnle}, {L{\'o}pez-Caniego},
  {Lubin}, {Mac{\'\i}as-P{\'e}rez}, {Maggio}, {Maino}, {Mandolesi}, {Mangilli},
  {Maris}, {Martin}, {Mart{\'\i}nez-Gonz{\'a}lez}, {Masi}, {Matarrese},
  {McGehee}, {Meinhold}, {Melchiorri}, {Mendes}, {Mennella}, {Migliaccio},
  {Mitra}, {Miville-Desch{\^e}nes}, {Molinari}, {Moneti}, {Montier},
  {Morgante}, {Mortlock}, {Moss}, {Munshi}, {Murphy}, {Naselsky}, {Nati},
  {Natoli}, {Netterfield}, {N{\o}rgaard-Nielsen}, {Noviello}, {Novikov},
  {Novikov}, {Oppermann}, {Oxborrow}, {Paci}, {Pagano}, {Pajot}, {Paoletti},
  {Pasian}, {Patanchon}, {Perdereau}, {Perotto}, {Perrotta}, {Pettorino},
  {Piacentini}, {Piat}, {Pierpaoli}, {Pietrobon}, {Plaszczynski},
  {Pointecouteau}, {Polenta}, {Popa}, {Pratt}, {Pr{\'e}zeau}, {Prunet},
  {Puget}, {Rachen}, {Rebolo}, {Reinecke}, {Remazeilles}, {Renault}, {Renzi},
  {Ristorcelli}, {Rocha}, {Rosset}, {Rossetti}, {Roudier},
  {Rubi{\~n}o-Mart{\'\i}n}, {Ruiz-Granados}, {Rusholme}, {Sandri}, {Santos},
  {Savelainen}, {Savini}, {Scott}, {Seiffert}, {Shellard}, {Shiraishi},
  {Spencer}, {Stolyarov}, {Stompor}, {Sudiwala}, {Sunyaev}, {Sutton},
  {Suur-Uski}, {Sygnet}, {Tauber}, {Terenzi}, {Toffolatti}, {Tomasi},
  {Tristram}, {Tucci}, {Tuovinen}, {Umana}, {Valenziano}, {Valiviita}, {Van
  Tent}, {Vielva}, {Villa}, {Wade}, {Wandelt}, {Wehus}, {Yvon}, {Zacchei}, \&
  {Zonca}}]{planck_2016}
{Planck Collaboration}, {Ade}, P.~A.~R., {Aghanim}, N., {et~al.} 2016, \aap,
  594, A19

\bibitem[{{Planck Collaboration} {et~al.}(2020){Planck Collaboration},
  {Aghanim}, {Akrami}, {Ashdown}, {Aumont}, {Baccigalupi}, {Ballardini},
  {Banday}, {Barreiro}, {Bartolo}, {Basak}, {Battye}, {Benabed}, {Bernard},
  {Bersanelli}, {Bielewicz}, {Bock}, {Bond}, {Borrill}, {Bouchet}, {Boulanger},
  {Bucher}, {Burigana}, {Butler}, {Calabrese}, {Cardoso}, {Carron},
  {Challinor}, {Chiang}, {Chluba}, {Colombo}, {Combet}, {Contreras}, {Crill},
  {Cuttaia}, {de Bernardis}, {de Zotti}, {Delabrouille}, {Delouis}, {Di
  Valentino}, {Diego}, {Dor{\'e}}, {Douspis}, {Ducout}, {Dupac}, {Dusini},
  {Efstathiou}, {Elsner}, {En{\ss}lin}, {Eriksen}, {Fantaye}, {Farhang},
  {Fergusson}, {Fernandez-Cobos}, {Finelli}, {Forastieri}, {Frailis},
  {Fraisse}, {Franceschi}, {Frolov}, {Galeotta}, {Galli}, {Ganga},
  {G{\'e}nova-Santos}, {Gerbino}, {Ghosh}, {Gonz{\'a}lez-Nuevo}, {G{\'o}rski},
  {Gratton}, {Gruppuso}, {Gudmundsson}, {Hamann}, {Handley}, {Hansen},
  {Herranz}, {Hildebrandt}, {Hivon}, {Huang}, {Jaffe}, {Jones}, {Karakci},
  {Keih{\"a}nen}, {Keskitalo}, {Kiiveri}, {Kim}, {Kisner}, {Knox},
  {Krachmalnicoff}, {Kunz}, {Kurki-Suonio}, {Lagache}, {Lamarre}, {Lasenby},
  {Lattanzi}, {Lawrence}, {Le Jeune}, {Lemos}, {Lesgourgues}, {Levrier},
  {Lewis}, {Liguori}, {Lilje}, {Lilley}, {Lindholm}, {L{\'o}pez-Caniego},
  {Lubin}, {Ma}, {Mac{\'\i}as-P{\'e}rez}, {Maggio}, {Maino}, {Mandolesi},
  {Mangilli}, {Marcos-Caballero}, {Maris}, {Martin}, {Martinelli},
  {Mart{\'\i}nez-Gonz{\'a}lez}, {Matarrese}, {Mauri}, {McEwen}, {Meinhold},
  {Melchiorri}, {Mennella}, {Migliaccio}, {Millea}, {Mitra},
  {Miville-Desch{\^e}nes}, {Molinari}, {Montier}, {Morgante}, {Moss}, {Natoli},
  {N{\o}rgaard-Nielsen}, {Pagano}, {Paoletti}, {Partridge}, {Patanchon},
  {Peiris}, {Perrotta}, {Pettorino}, {Piacentini}, {Polastri}, {Polenta},
  {Puget}, {Rachen}, {Reinecke}, {Remazeilles}, {Renzi}, {Rocha}, {Rosset},
  {Roudier}, {Rubi{\~n}o-Mart{\'\i}n}, {Ruiz-Granados}, {Salvati}, {Sandri},
  {Savelainen}, {Scott}, {Shellard}, {Sirignano}, {Sirri}, {Spencer},
  {Sunyaev}, {Suur-Uski}, {Tauber}, {Tavagnacco}, {Tenti}, {Toffolatti},
  {Tomasi}, {Trombetti}, {Valenziano}, {Valiviita}, {Van Tent}, {Vibert},
  {Vielva}, {Villa}, {Vittorio}, {Wandelt}, {Wehus}, {White}, {White},
  {Zacchei}, \& {Zonca}}]{planck_2020_cosmology2018}
{Planck Collaboration}, {Aghanim}, N., {Akrami}, Y., {et~al.} 2020, \aap, 641,
  A6

\bibitem[{{Press} \& {Schechter}(1974)}]{press_schechter_model_1974}
{Press}, W.~H. \& {Schechter}, P. 1974, \apj, 187, 425

\bibitem[{{Quashnock} {et~al.}(1989){Quashnock}, {Loeb}, \&
  {Spergel}}]{quashnock_1989_QCD_bfields}
{Quashnock}, J.~M., {Loeb}, A., \& {Spergel}, D.~N. 1989, \apjl, 344, L49

\bibitem[{Rincon(2019)}]{rincon_2019_dynamothies}
Rincon, F. 2019, Journal of Plasma Physics, 85, 205850401

\bibitem[{{Rincon} {et~al.}(2016){Rincon}, {Califano}, {Schekochihin}, \&
  {Valentini}}]{Rincon_2016_colless_dynamo}
{Rincon}, F., {Califano}, F., {Schekochihin}, A.~A., \& {Valentini}, F. 2016,
  Proceedings of the National Academy of Science, 113, 3950

\bibitem[{{Riquelme} {et~al.}(2015){Riquelme}, {Quataert}, \&
  {Verscharen}}]{riquelme_quataert_verscharen_2105}
{Riquelme}, M.~A., {Quataert}, E., \& {Verscharen}, D. 2015, \apj, 800, 27

\bibitem[{{Rudakov} \& {Sagdeev}(1961)}]{vedenov_sagdeev_1958}
{Rudakov}, L.~I. \& {Sagdeev}, R.~Z. 1961, Soviet Physics Doklady, 6, 415

\bibitem[{{Ryu} {et~al.}(1998){Ryu}, {Kang}, \& {Biermann}}]{biermann_ryu_1998}
{Ryu}, D., {Kang}, H., \& {Biermann}, P.~L. 1998, \aap, 335, 19

\bibitem[{{Santos-Lima} {et~al.}(2014){Santos-Lima}, {de Gouveia Dal Pino},
  {Kowal}, {Falceta-Gon{\c{c}}alves}, {Lazarian}, \&
  {Nakwacki}}]{santos-lima-2014}
{Santos-Lima}, R., {de Gouveia Dal Pino}, E.~M., {Kowal}, G., {et~al.} 2014,
  \apj, 781, 84

\bibitem[{{Schekochihin} \& {Cowley}(2006)}]{Schekochihin_2006_magfield_gclust}
{Schekochihin}, A.~A. \& {Cowley}, S.~C. 2006, Physics of Plasmas, 13, 056501

\bibitem[{{Schekochihin} {et~al.}(2002){Schekochihin}, {Cowley}, {Hammett},
  {Maron}, \& {McWilliams}}]{scheko_2002_nonlin_MHD_dnyamo}
{Schekochihin}, A.~A., {Cowley}, S.~C., {Hammett}, G.~W., {Maron}, J.~L., \&
  {McWilliams}, J.~C. 2002, New Journal of Physics, 4, 84

\bibitem[{Schekochihin {et~al.}(2004)Schekochihin, Cowley, Taylor, Maron, \&
  McWilliams}]{Schekochihin_2004_small_scale_dynamo}
Schekochihin, A.~A., Cowley, S.~C., Taylor, S.~F., Maron, J.~L., \& McWilliams,
  J.~C. 2004, The Astrophysical Journal, 612, 276

\bibitem[{{Scherrer} {et~al.}(1977){Scherrer}, {Wilcox}, {Svalgaard}, {Duvall},
  {Dittmer}, \& {Gustafson}}]{Sun_mean_magfield_Scherrer1977}
{Scherrer}, P.~H., {Wilcox}, J.~M., {Svalgaard}, L., {et~al.} 1977, Solar
  Physics, 54, 353

\bibitem[{{Schleicher} {et~al.}(2013){Schleicher}, {Schober}, {Federrath},
  {Bovino}, \& {Schmidt}}]{schleicher_2013_ssd}
{Schleicher}, D. R.~G., {Schober}, J., {Federrath}, C., {Bovino}, S., \&
  {Schmidt}, W. 2013, New Journal of Physics, 15, 023017

\bibitem[{{Schmassmann} {et~al.}(2018){Schmassmann}, {Schlichenmaier}, \&
  {Bello Gonz{\'a}lez}}]{sunspots_magfields}
{Schmassmann}, M., {Schlichenmaier}, R., \& {Bello Gonz{\'a}lez}, N. 2018,
  \aap, 620, A104

\bibitem[{{Schober} {et~al.}(2022){Schober}, {Rogachevskii}, \&
  {Brandenburg}}]{schober_2022_chiral_anomaly}
{Schober}, J., {Rogachevskii}, I., \& {Brandenburg}, A. 2022, \prl, 128, 065002

\bibitem[{{Schober} {et~al.}(2013){Schober}, {Schleicher}, \&
  {Klessen}}]{schober_2013_youngal}
{Schober}, J., {Schleicher}, D.~R.~G., \& {Klessen}, R.~S. 2013, \aap, 560, A87

\bibitem[{{Sharma} {et~al.}(2006){Sharma}, {Hammett}, {Quataert}, \&
  {Stone}}]{Sharma_2006_hwl}
{Sharma}, P., {Hammett}, G.~W., {Quataert}, E., \& {Stone}, J.~M. 2006, \apj,
  637, 952

\bibitem[{{Shi} {et~al.}(2018){Shi}, {Nagai}, \& {Lau}}]{shi_2018}
{Shi}, X., {Nagai}, D., \& {Lau}, E.~T. 2018, \mnras, 481, 1075

\bibitem[{{Snyder} {et~al.}(1997){Snyder}, {Hammett}, \&
  {Dorland}}]{snyder_1997}
{Snyder}, P.~B., {Hammett}, G.~W., \& {Dorland}, W. 1997, Physics of Plasmas,
  4, 3974

\bibitem[{{Southwood} \& {Kivelson}(1993)}]{southwood_kivelson_1993}
{Southwood}, D.~J. \& {Kivelson}, M.~G. 1993, \jgr, 98, 9181

\bibitem[{{Springel} {et~al.}(2005){Springel}, {White}, {Jenkins}, {Frenk},
  {Yoshida}, {Gao}, {Navarro}, {Thacker}, {Croton}, {Helly}, {Peacock}, {Cole},
  {Thomas}, {Couchman}, {Evrard}, {Colberg}, \&
  {Pearce}}]{springel_2005_millenium}
{Springel}, V., {White}, S. D.~M., {Jenkins}, A., {et~al.} 2005, \nat, 435, 629

\bibitem[{{St-Onge}(2019)}]{st-onge_thesis}
{St-Onge}, D.~A. 2019, arXiv e-prints, arXiv:1912.11072

\bibitem[{{St-Onge} \& {Kunz}(2018)}]{st-onge_2018_hybrid}
{St-Onge}, D.~A. \& {Kunz}, M.~W. 2018, \apjl, 863, L25

\bibitem[{{St-Onge} {et~al.}(2020){St-Onge}, {Kunz}, {Squire}, \&
  {Schekochihin}}]{st-onge_kunz_squire_scheko_20202}
{St-Onge}, D.~A., {Kunz}, M.~W., {Squire}, J., \& {Schekochihin}, A.~A. 2020,
  Journal of Plasma Physics, 86, 905860503

\bibitem[{{Subramanian} {et~al.}(1994){Subramanian}, {Narasimha}, \&
  {Chitre}}]{biermann_subramanian_1994}
{Subramanian}, K., {Narasimha}, D., \& {Chitre}, S.~M. 1994, \mnras, 271, L15

\bibitem[{{Subramanian} {et~al.}(2006){Subramanian}, {Shukurov}, \&
  {Haugen}}]{subramanian_2006}
{Subramanian}, K., {Shukurov}, A., \& {Haugen}, N. E.~L. 2006, \mnras, 366,
  1437

\bibitem[{{Talebian} {et~al.}(2020){Talebian}, {Nassiri-Rad}, \&
  {Firouzjahi}}]{talebian_2020_infl_magnetogenesis}
{Talebian}, A., {Nassiri-Rad}, A., \& {Firouzjahi}, H. 2020, \prd, 102, 103508

\bibitem[{{T{\"o}rnkvist}(1998)}]{tornkvist_1998_electroweak_bfield}
{T{\"o}rnkvist}, O. 1998, \prd, 58, 043501

\bibitem[{{Tweed} {et~al.}(2009){Tweed}, {Devriendt}, {Blaizot}, {Colombi}, \&
  {Slyz}}]{tweed_2009_mt_extract_from_sim}
{Tweed}, D., {Devriendt}, J., {Blaizot}, J., {Colombi}, S., \& {Slyz}, A. 2009,
  \aap, 506, 647

\bibitem[{{Vazza} {et~al.}(2012){Vazza}, {Roediger}, \&
  {Br{\"u}ggen}}]{vazza_roediger_2012}
{Vazza}, F., {Roediger}, E., \& {Br{\"u}ggen}, M. 2012, \aap, 544, A103

\bibitem[{{Vogt} \& {En{\ss}lin}(2003)}]{vogt_2003_abell_400_2634_hydra}
{Vogt}, C. \& {En{\ss}lin}, T.~A. 2003, \aap, 412, 373

\bibitem[{{Wallbank} {et~al.}(2022){Wallbank}, {Maughan}, {Gastaldello},
  {Potter}, \& {Wik}}]{wallbank_2022_xray_obs_temp}
{Wallbank}, A.~N., {Maughan}, B.~J., {Gastaldello}, F., {Potter}, C., \& {Wik},
  D.~R. 2022, \mnras, 517, 5594

\bibitem[{{Widrow}(2002)}]{widrow_2002_gal_bf}
{Widrow}, L.~M. 2002, Reviews of Modern Physics, 74, 775

\bibitem[{{Wright}(2006)}]{Wright2006_cosmology_calculator}
{Wright}, E.~L. 2006, \pasp, 118, 1711

\end{thebibliography}

\begin{acknowledgements}
We are very grateful
to Matthew Kunz and Jonathan Squire for providing us with all the necessary ingredients for modeling magnetic fields in the intracluster medium, during the Nordita program ``Magnetic Field Evolution in Low Density or Strongly Stratified Plasmas'' in 2022. 
Undoubtedly, our work would not have been possible without their contribution to the physics of weakly collisional plasmas.
We further thank Abhijit B.\ Bendre for his numerous comments on this study
and the referee for a very constructive report.
Finally, we acknowledge the support from the Swiss National Science Foundation under Grant No.\ 185863.
\end{acknowledgements}

\begin{appendix}

\section{GALFORM \& Modified GALFORM models}\label{appendix:galform}
The $\textsc{GALFORM}$ model \citep{cole2000_galform} is based 
upon the extended Press-Schechter theory \citep{press_schechter_model_1974}, which considers 
the following conditional mass function:
\begin{equation}\label{eq:mass_function}
\begin{split}
f(M_1|M_2)d\ln M_1 = &\sqrt{\frac{2}{\pi}}\frac{\sigma_1^2(\delta_1-\delta_2)}{(\sigma^2-\sigma_2^2)^{3/2}}\\
&\exp\left(-\frac{1}{2}\frac{(\delta_1-\delta_2)^2}{\sigma_1^2-\sigma_2^2}\right) \left| \frac{d\ln \sigma}{d \ln M_1}\right|d\ln M_1
\end{split}
\end{equation}
where $f(M_1|M_2)$ represents the fraction of mass from halos 
of mass $M_2$ at redshift $z_2$ that is contained in progenitor 
halos of mass $M_1$ at earlier redshift $z_1$.
The parameter $\delta_i$ denotes the linear overdensity 
threshold for spherical collapse at redshift $z_i$. 
The quantity $\sigma(M)$ denotes the rms linear density fluctuation extrapolated at $z=0$ in spheres containing mass $M$. For simplicity, we adopt $\sigma_i(M)\equiv \sigma_i$. Taking the limit $z_1 \to z_2$, we get 
\begin{equation}
\begin{split}
&\left. \frac{df}{dz_1} \right|_{z_1=z_2}d\ln M_1 dz_1\\
& = \sqrt{\frac{2}{\pi}}\frac{\sigma_1^2}{(\sigma_1^2-\sigma_2^2)^{3/2}} \frac{d\delta_1}{dz_1}\left| \frac{d\ln \sigma_1}{d \ln M_1}\right|d \ln M_1 dz_1
\end{split}
\end{equation}
which allows obtaining the mean number of 
halos 
of mass $M_1$ into which a halo of mass $M_2$ splits when we increase redshift upwards with a redshift step $dz_1$:
\begin{equation}\label{eq:frac_halos}
\frac{dN}{dM_1} = \frac{1}{M_1}\frac{df}{dz_1}\frac{M_2}{M_1}dz_1.
\end{equation}
Assuming a resolution mass $M_{\mathrm{res}}$, we can also determine the average number of progenitors with masses $M_1$ in the interval $M_{\mathrm{res}}< M_1 < M_2/2$, which is given by 
\begin{equation}\label{eq:PS_prob_halos}
P = \int_{M_{\mathrm{res}}}^{M_2/2} \frac{dN}{dM_1}dM_1,
\end{equation}
along with the fraction of mass of the final object in progenitors below $M_{\mathrm{res}}$, given by 
\begin{equation}
F = \int_0^{M_{\mathrm{res}}}\frac{dN}{dM_1}\frac{M_1}{M_2}dM_1.
\end{equation}
The GALFORM algorithm operates as follows. We start with the redshift and the mass of the final halo. Then we consider a redshift step $dz_1$ that satisfies $P \ll 1$ (meaning that a halo is unlikely to have more than two progenitors at redshift $z+dz$). A random number $R\in [0,1]$ is generated. If $R>P$ the main halo is not split, and its mass is reduced to $M_2(1-F)$ to account for the accreted mass. 
If $R \leq P$, then a random mass $M_{\mathrm{res}} > M_1 > M_2/2$ is generated to produce two 
new halos with masses $M_1$ and $M_2(1-F)-M_1$. 
This same process is repeated until the maximum redshift or the resolution mass is reached.  

However, there are some issues with such an approach. One of them is that the conditional mass function \eqref{eq:mass_function} does not match what is found in $N$-body simulations. However, statistical properties produced by the GALFORM algorithm have similar trends with those of merger trees constructed from high-resolution $N$-body simulations, with an error on mass progenitors that increases with redshift. In that spirit, \citet{parkinson_2008_mod_galform} replaced the initial statistics Eq. \eqref{eq:frac_halos} with
\begin{equation}
\frac{dN}{dM_1} \to \frac{dN}{dM_1}G(\sigma_1/\sigma_2, \sigma_2/\sigma_2),
\end{equation}
where $G$ is a ``perturbative''
function that matches the generated merger tree with simulations. They also made the following assumption
\begin{equation}
G(\sigma_1/\sigma_2, \sigma_2/\sigma_2) = G_0\left(\frac{\sigma_1}{\sigma_2}\right)^{\gamma_1}\left(\frac{\delta_2}{\sigma_2}\right)^{\gamma_2}
\end{equation}
for which they obtained $G_0 = 0.57, \gamma_1 = 0.38$ and $\gamma_2=-0.01$. Finally, they obtained halo abundances consistent with the Sheth-Tormen mass function \citep[see][and references therein]{parkinson_2008_mod_galform}. In our work, we use the same values for $G_0, \gamma_1$, and $\gamma_2$.

% **********************************************************************************************

\section{Effect number of generated trees}\label{appendix:nb_trees_effect}
\begin{figure}
    \centering
    \includegraphics[width = 0.9\columnwidth]{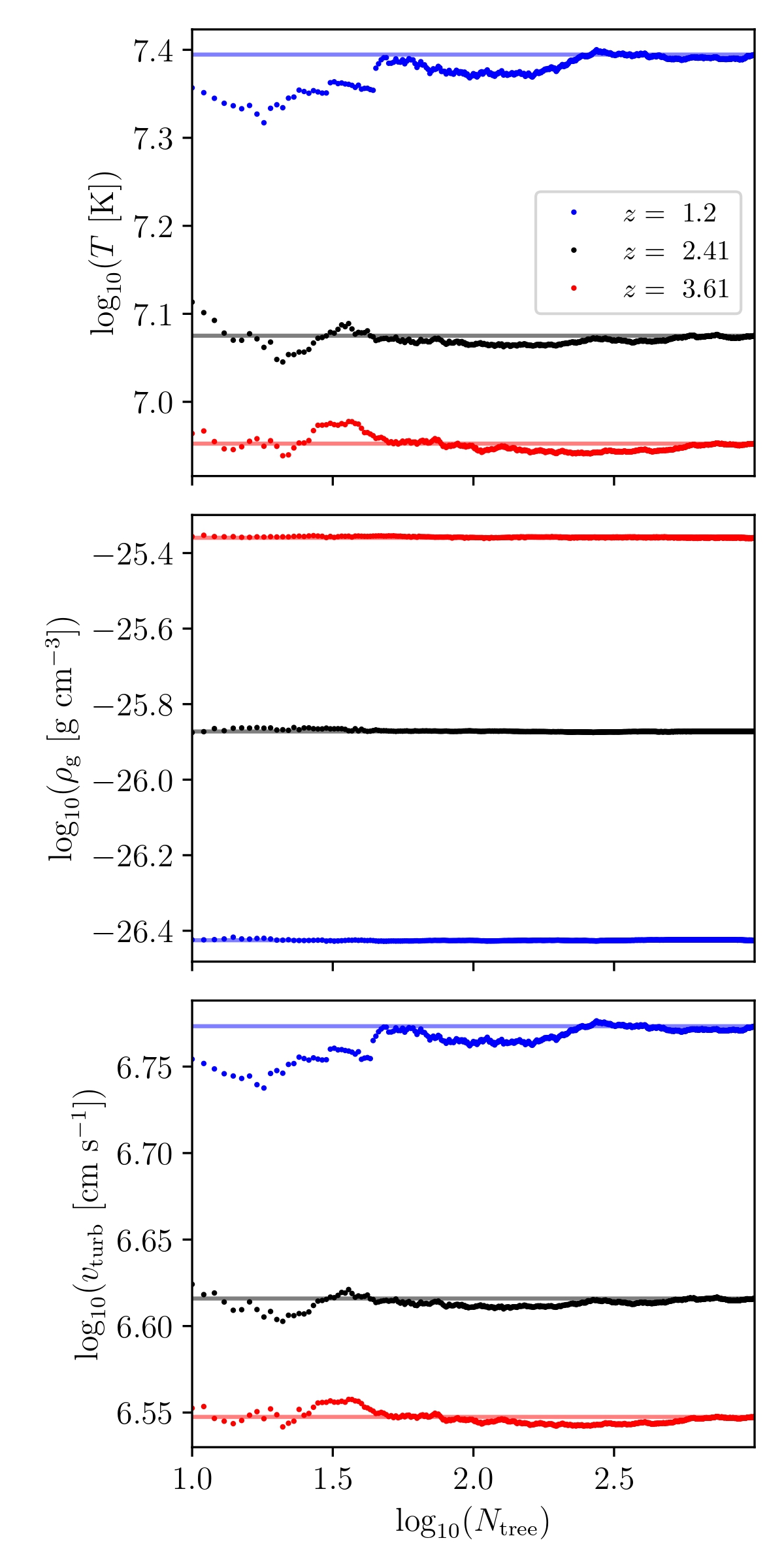}
    \caption{Skew-normal average of the temperature, gas density, and turbulent velocity at three different values of redshift, as a function of the number of generated merger trees. The solid lines represent the values of the physical quantities for $N_{\mathrm{tree}} = 10^3$.}
    \label{fig:nb_trees_effect}
\end{figure}

Figure~\ref{fig:nb_trees_effect} shows the effect of the number of generated merger trees, on the Skew-normal average (Sec.~\ref{appendix:skew-normal}) calculated for the temperature, the baryonic gas density, and the turbulent velocity, at three different redshifts. 
The solid horizontal lines correspond to the value calculated with $N_{\mathrm{tree}} = 10^3$ merger trees. 
It is evident that convergence is reached, even for $N_{\mathrm{tree}} \gtrsim 10^{2.5}$ merger trees. 
This justifies our choice of setting $N_{\mathrm{tree}} = 10^3$.
% **********************************************************************************************
\section{Skew-normal averaging process}\label{appendix:skew-normal}

Here we present our method for averaging the time evolution curves of all $N_{\mathrm{tree}} = 10^3$ merger trees, for all physical quantities. 
Initially, one might consider using the conventional arithmetic average to estimate the overall behavior of these curves at each specific redshift. 
However, this approach presents a significant flaw when analyzing the evolution of the magnetic field. By examining the left panel of Fig.~\ref{fig:bfield_evolution_example_with_error}, it becomes evident that at approximately $z \simeq 0.75$, while a few magnetic field curves have reached equipartition, the majority are still in the phase of exponential growth, with values 
between $B \approx 10^{-14}$ G and $B \approx 10^{-12}$ G. 
Consequently, applying the classical arithmetic average to this dataset would yield a value close to equipartition, which does not accurately represent the actual trend exhibited by all the curves.

\begin{figure}
    \centering
    \includegraphics[width = 0.8\columnwidth]{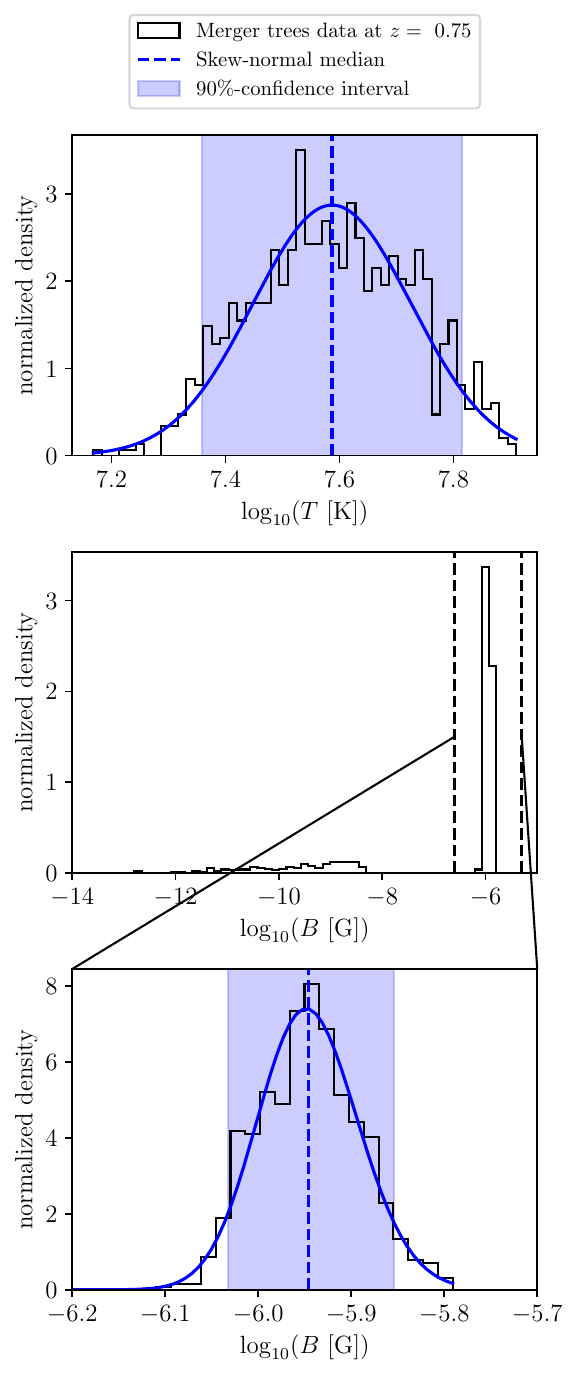}
    \caption{Illustration of the method using a Skew-normal distribution to average different physical quantities at a given redshift, for many merger trees.
    (top) Distribution of the temperature values from all merger trees at redshift $z=0.75$.
    The blue line represents the Skew-normal distribution used to fit the data. The dashed line and the blue-shaded respectively are the distribution's median (that is used as an estimator for the average of all curves), and its corresponding 90\%-confidence interval. (middle) Distribution of the magnetic field values at this same redshift. The two dashed black lines indicate the region that is fitted with the Skew-normal distribution. (bottom) Fitting of the region in the middle panel.
    Those three panels were computed with $\mathrm{Re}_{\mathrm{eff}}$ Model L, for $\log_{10}(B_0~[\mathrm{G}]) = -20$, $\alpha_0 = 20$, and $z_{\mathrm{start}} = 1.5$.
    }
    \label{fig:skew_normal_avg}
\end{figure}

We decided to adopt the following strategy.
At each given redshift, we fit the distribution of the values of all curves with a 
Skew-normal distribution.
However, special attention has to be paid to the magnetic 
field, as it is shown in Fig.~\ref{fig:skew_normal_avg}.
Here, we show the histograms of different quantities, 
where the $y$-axis is normalized 
so that the total area of the histogram equals one.
The top panel shows the temperature distribution at redshift $z = 0.29$, and the middle one the distribution of the magnetic field at the same redshift. If we were to fit the whole magnetic field sample, a skew-normal distribution would not necessarily be suitable as two distinct maxima appear. Therefore, we adopted the following convention. 
Whenever the number of trees that have reached equipartition is dominant, we fit the region delimited by dashed lines on the middle panel, which is shown on the bottom panel. However, if this number is smaller than the trees outside this region, we fit the trees in the latter.    
Finally, let us note that the situation where the two histograms, distinguishing the equipartition magnetic field from other values, have the same number of occurrences is extremely rare.
% **********************************************************************************************
\section{Time-resolution of the dynamo model}\label{appendix:time_res}

\begin{figure*}
    \centering
    \includegraphics[width = \textwidth]{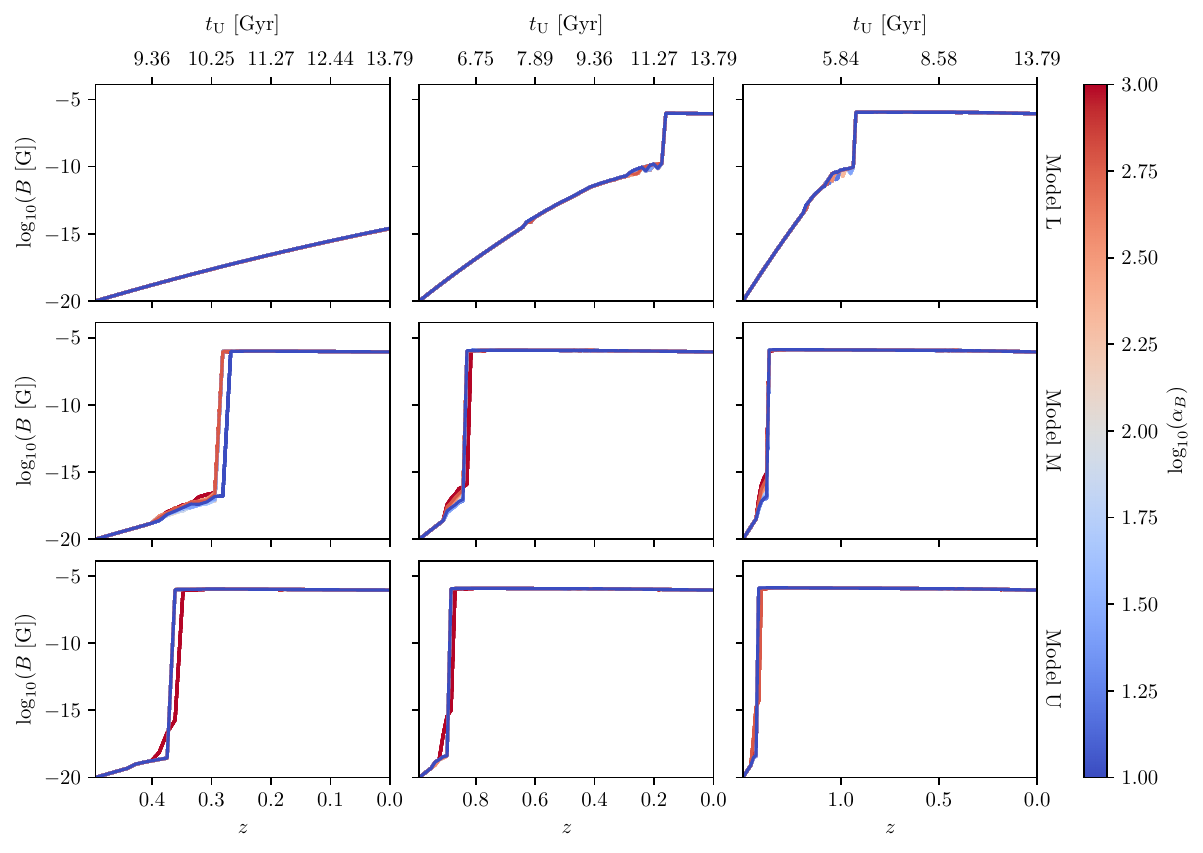}
    \caption{Magnetic field evolution for different values of $\alpha_B$. Each column corresponds to a different value of redshift at which the dynamo starts. Each row corresponds to a different model for the effective Reynolds number described in Sec. \ref{sec:model_magfields}. }
    \label{fig:alphaB_calculation}
\end{figure*}

As described in Sec.~\ref{sec:model_magfields}, we solve Eq.~\eqref{eq:turbulent_dynamo} between every redshift step of our merger tree (we recall that $N_z = 300$). 
Redshift is converted to time using the cosmology calculator of
\citet{Wright2006_cosmology_calculator} with the cosmological parameters given in Sec.~\ref{subsec:general}.
However, considering only the two values of time between two 
redshift steps is not efficient enough to highlight the 
complexity of the different stages of the dynamo, like non-linear 
effects when the magnetic field is getting close to equipartition. 
Therefore, we create a discretized time grid, between two given 
redshift steps, whose time step is kept constant (meaning that 
the number of time grid cells between two different consecutive reshifts will vary). The constant time step is given as follows. We can see that Eq. \eqref{eq:turbulent_dynamo} is of the form $B(t) \propto e^{\Gamma_{\mathrm{eff}} t}$, where $\Gamma_{\mathrm{eff}} \equiv (v_{\mathrm{turb}}/L_0)\mathrm{Re}_{\mathrm{eff}}^{1/2}$ is the growth rate of the magnetic field. 
However, given our construction for the effective 
Reynolds number in Sec.\ \ref{sec:model_magfields}, it
is obvious that $\mathrm{Re}_{\mathrm{eff}} \geq \mathrm{Re}$, which would yield a smaller timescale of the growth rate than with the classical 
Reynolds number. 
Therefore, in order to avoid over-discretizing our grid, we consider the timescale coming from the classical 
Reynolds number,
$\Gamma^{-1}$, to be a minimum timescale that should not be exceeded. Therefore, the constant value $\Delta t$ of all grid cells $N_{\Delta t}$ contained in our discretized time between two redshift steps is given by 
\begin{equation}\label{eq:time_res_bfield}
\Delta t \equiv \frac{\Gamma^{-1}}{\alpha_B} = \frac{1}{\alpha_B}\left(\frac{v_{\mathrm{turb}}}{L_0} \mathrm{Re}^{1/2} \right)^{-1},
\end{equation}
where $\alpha_B$ is a constant value. 

We determine the latter by plotting the evolution of the magnetic field, for a fixed set of merger-tree parameters, and for different values of $\alpha_B$. Fig.\ \ref{fig:alphaB_calculation} shows such results for $B_0 = 10^{-15}~\mathrm{G}$ and $\alpha_0 = 20$. It appears that when $\alpha_B \approx 10^{3.5}$, all
curves converge. Therefore, in our work, we adopt a constant value of $\alpha_B = 10^{4}$. Note that we have checked that this value of $\alpha_B$ does not produce other results for all other parameters of $B_0$ and $\alpha_0$.

\section{Estimation of the $c$-parameter}\label{appendix:c_param}
We describe here our method to determine the concentration parameter in every subhalo of the generated merger trees. We start by initiating a value of $c$ for 
subhalos 
at $z_{\mathrm{max}}$, with
\begin{equation}\label{eq:init_c_values}
c(M) = 4.67\left(\frac{M}{10^{14}h^{-1}M_{\odot}}\right)^{-0.11},
\end{equation}
which was proposed by \citet{neto_2007_cold_dm_statistics}, where $h$ is the Hubble constant in units of $100~\mathrm{km}~\mathrm{s}^{-1}~\mathrm{Mpc}^{-1}$ \citep[see][]{springel_2005_millenium}.
This choice of the initial condition given in Eq.~\ref{eq:init_c_values} 
is somehow arbitrary, but after trying different values of $c$ for the 
earliest halos, we have found that the final result does not depend 
on this particular choice. 
We calculate the total energy of every subhalo using the same expressions adopted in \citet{Johnson_2021_random_walk_c}, which is the sum of its kinetic and potential energy, which are respectively given by
\begin{equation}\label{eq:halos_kinetic_enregy}
T = 2\pi \left[r_{\mathrm{vir}}^3\rho(r_{\mathrm{vir}})\sigma^2(r_{\mathrm{vir}})+\int_{0}^{r_{\mathrm{vir}}}GM(r)\rho(r)
r \mathrm{d}r\right]
\end{equation}
and 
\begin{equation}\label{eq:halos_grav_enregy}
W = -\frac{G}{2}\left[\int_0^{r_{\mathrm{vir}}}\frac{M^2(r)}{r^2}
\mathrm{d}r+
\frac{M^2(r_{\mathrm{vir}})}{r_{\mathrm{vir}}}\right].
\end{equation}
Here, $\sigma(r)$ is the velocity dispersion that is determined by the Jeans equation:
\begin{equation}
\frac{d(\rho \sigma^2)}{dr} = -\rho(r)\frac{GM(r)}{r^2}.
\end{equation}

Therefore, each halo is attributed a total energy of $E_\mathrm{h} = T+W$, which depends on parameters 
$\rho_\mathrm{s}$ 
and $r_\mathrm{s}$. 
When a subhalo is the result of $N$ merging subhalos, 
we compute the energy 
$E_\mathrm{h,f} = T_\mathrm{f}+W_\mathrm{f}$ 
of the resulting halo, which we equalize with the energy of the system of all progenitors. We express the total energy $E_{N}$ of a system of $N$ subhalos using the same strategy presented in \citet{dvorkin_Rephaeli_2011}, which is
\begin{equation}\label{eq:total_energy_subhalos}
E_N = \sum_{i=1}^{N}E_{\mathrm{h},i}+\sum_{i\neq j}U_{ij}+\sum_{i\neq j}U_{\mathrm{acc}, ij}
\end{equation}
where 
\begin{equation}
 U_{ij} \equiv -\frac{GM_iM_j}{d_{0,ij}}   
\end{equation}
is the gravitational energy of two halos at the largest point of their separation, with $d_{0, ij}$ being the distance at which subhalos $i$ and $j$ are bound, and where 
\begin{equation}
U_{\mathrm{acc}, ij} \equiv - \frac{G(M_i+M_j)\Delta M }{r_\mathrm{vir,f}}
\end{equation}
is the energy component of the accreted matter, where  
$r_\mathrm{vir,f}$ stands 
for the virial radius of the resulting halo, and where 
$\Delta M \equiv M_\mathrm{f} - \sum_{i}M_i$ 
is the mass that was not resolved during the merging event, below the value of $M_{\mathrm{res}}$. We use the same value of $d_0$ as the one adopted in \citet{dvorkin_Rephaeli_2011}, which is 
$d_{0, ij} = 5(r_{\mathrm{vir},i}+r_{\mathrm{vir},j})$. 
Our strategy is then to compute the energy for each subhalo (whether it has one or multiple progenitors) and find the value of $c$ that matches the energy conservation with its progenitors. Once $c$ is calculated, the parameter $\rho_\mathrm{s}$ is obtained by integration of the dark matter density to match the mass of each subhalo.

\end{appendix}
\end{document}